\documentclass[sigplan,screen]{acmart}

\AtBeginDocument{%
  \providecommand\BibTeX{{%
    \normalfont B\kern-0.5em{\scshape i\kern-0.25em b}\kern-0.8em\TeX}}}

\copyrightyear{2020} 
\acmYear{2020} 
\setcopyright{acmcopyright}
\acmConference[ASPLOS '20]{Proceedings of the Twenty-Fifth International Conference on Architectural Support for Programming Languages and Operating Systems}{March 16--20, 2020}{Lausanne, Switzerland}
\acmBooktitle{Proceedings of the Twenty-Fifth International Conference on Architectural Support for Programming Languages and Operating Systems (ASPLOS '20), March 16--20, 2020, Lausanne, Switzerland}
\acmPrice{15.00}
\acmDOI{10.1145/3373376.3378469}
\acmISBN{978-1-4503-7102-5/20/03}

\usepackage[vlined,linesnumbered]{algorithm2e}
\usepackage{amsthm}

\usepackage{thmtools}
\declaretheoremstyle[%
  spaceabove=3pt,%
  spacebelow=3pt,%
  headfont=\normalfont\itshape,%
  postheadspace=1em,%
  qed=\qedsymbol%
]{mystyle} 
\declaretheorem[name={Proof},style=mystyle,unnumbered,
]{prf}

\newtheoremstyle{exampstyle}
  {1pt} % Space above
  {1pt} % Space below
  {} % Body font
  {} % Indent amount
  {\bfseries} % Theorem head font
  {.} % Punctuation after theorem head
  {.5em} % Space after theorem head
  {} % Theorem head spec (can be left empty, meaning `normal')

\theoremstyle{exampstyle}

\theoremstyle{exampstyle} \newtheorem{theorem}{Theorem}[section]
\newtheorem{lemma}[theorem]{Lemma}
\usepackage{xspace} % TODO: should be removed

\usepackage{hyperref}

\usepackage{multirow} % multi-column
\usepackage{dcolumn} % digit alignment in column
\usepackage{bm} % bold font in math mode
\usepackage{subcaption} % TODO: should be removed
\usepackage{listings}

\definecolor{mygreen}{HTML}{C6D645}
\definecolor{myyellow}{HTML}{F4D54E}
\definecolor{myred}{HTML}{FC4C01}

\newcommand{\Dbegin}{D\!.\mathit{begin}}
\newcommand{\Dend}{D\!.\mathit{end}}
\newcommand{\Cbegin}{C\!.\mathit{begin}}
\newcommand{\Cend}{C\!.\mathit{end}}

\newcommand{\MemOperand}{\texttt{<mem>}\xspace}
\newcommand{\ImmdOperand}{\texttt{<imm>}\xspace}
\newcommand{\RegOperand}{\texttt{<reg>}\xspace}
\newcommand{\ScratchReg}{\texttt{<scratch\_reg>}\xspace}
\newcommand{\DomainId}{\texttt{<domain\_id>}\xspace}

\newcommand{\MemGuard}{\texttt{mem\_guard}\xspace}
\newcommand{\CfiGuard}{\texttt{cfi\_guard}\xspace}
\newcommand{\CfiLabel}{\texttt{cfi\_label}\xspace}

\newcommand{\MemGuards}{\texttt{mem\_guards}\xspace}
\newcommand{\CfiGuards}{\texttt{cfi\_guards}\xspace}
\newcommand{\CfiLabels}{\texttt{cfi\_labels}\xspace}

\newcommand{\Bnd}[1]{\texttt{bnd#1}\xspace}

\begin{document}

% Clear fancy headers for the camera-ready version
% ACM is likely to add headers for their published version
\fancyhead{}

\title{Occlum: Secure and Efficient Multitasking Inside a Single Enclave of Intel SGX}

\author{Youren Shen}
\authornote{Both authors contributed equally to this research.}
\email{syr15@mails.tsinghua.edu.cn}
\affiliation{%
  \institution{Tsinghua University}
}

\author{Hongliang Tian}
\authornotemark[1]
\email{tate.thl@antfin.com}
\affiliation{%
  \institution{Ant Financial Services Group}
}

\author{Yu Chen}
\email{yuchen@mail.tsinghua.edu.cn}
\affiliation{%
  \institution{Tsinghua University}
}
\affiliation{%
  \institution{Peng Cheng Laboratory}
}

\author{Kang Chen}
\email{chenkang@tsinghua.edu.cn}
\affiliation{%
  \institution{Tsinghua University}
}
\authornote{This is the corresponding author.}

\author{Runji Wang}
\email{wrj15@mails.tsinghua.edu.cn}
\affiliation{%
  \institution{Tsinghua University}
}
\affiliation{%
  \institution{Ant Financial Services Group}
}

% \additionalaffiliation{%
%   \institution{Ant Financial Services Group}
%   \city{Beijing}
%   \country{China}
% }

\author{Yi Xu}
\authornote{This work was done while the author was at Tsinghua University.}
\email{xu1369@purdue.edu}
\affiliation{%
  \institution{Purdue University}
}
\affiliation{%
  \institution{Ant Financial Services Group}
}

\author{Yubin Xia}
\email{xiayubin@sjtu.edu.cn}
\affiliation{%
  \institution{Shanghai Jiao Tong University}
}

\author{Shoumeng Yan}
\email{shoumeng.ysm@antfin.com}
\affiliation{%
  \institution{Ant Financial Services Group}
}

\renewcommand{\shortauthors}{Youren Shen and Hongliang Tian, et al.}

\begin{abstract}
Intel Software Guard Extensions (SGX) enables user-level code to create private memory regions called \emph{enclaves}, whose code and data are protected by the CPU from software and hardware attacks outside the enclaves. Recent work introduces library operating systems (LibOSes) to SGX so that legacy applications can run inside enclaves with few or even no modifications. As virtually any non-trivial application demands multiple processes, it is essential for LibOSes to support multitasking. However, none of the existing SGX LibOSes support multitasking both \emph{securely} and \emph{efficiently}.

This paper presents Occlum, a system that enables secure and efficient multitasking on SGX. We implement the LibOS processes as \emph{SFI-Isolated Processes (SIPs)}. SFI is a software instrumentation technique for sandboxing untrusted modules (called domains). We design a novel SFI scheme named \emph{MPX-based, Multi-Domain SFI (MMDSFI)} and leverage MMDSFI to enforce the isolation of SIPs. We also design an independent verifier to ensure the security guarantees of MMDSFI. With SIPs safely sharing the single address space of an enclave, the LibOS can implement multitasking efficiently. The Occlum LibOS outperforms the state-of-the-art SGX LibOS on multitasking-heavy workloads by up to $6,600\times$ on micro-benchmarks and up to $500\times$ on application benchmarks.
\end{abstract}

\begin{CCSXML}
<ccs2012>
   <concept>
       <concept_id>10002978.10003006.10003007.10003009</concept_id>
       <concept_desc>Security and privacy~Trusted computing</concept_desc>
       <concept_significance>500</concept_significance>
   </concept>
   <concept>
       <concept_id>10011007.10010940.10010941.10010949.10010957.10010959</concept_id>
       <concept_desc>Software and its engineering~Multiprocessing / multiprogramming / multitasking</concept_desc>
       <concept_significance>500</concept_significance>
    </concept>
 </ccs2012>
\end{CCSXML}

\ccsdesc[500]{Security and privacy~Trusted computing}
\ccsdesc[500]{Software and its engineering~Multiprocessing / multiprogramming / multitasking}

\keywords{Intel SGX, library OS, multitasking, Software Fault Isolation, Intel MPX}

\maketitle

\section{Introduction}\label{section:introduction}

Intel Software Guard Extensions (SGX)~\cite{IntelSGX} is a promising trusted execution environment (TEE) technology. It enables user-level code to create private memory regions called \emph{enclaves}, whose code and data are protected by the CPU from software attacks (e.g., a malicious OS) and hardware attacks (e.g., memory bus snooping) outside the enclaves. SGX provides a practical solution to the long-standing problem of secure computation on untrusted platforms such as public clouds. SGX developers who use Intel SGX SDK~\cite{SGXSDK} are required to partition SGX-protected applications into enclave and non-enclave halves. This leads to tremendous effort to refactor legacy code for SGX. Recent work~\cite{Haven, Scone,GrapheneSGX} tries to minimize the effort by introducing library operating systems (LibOSes)~\cite{Exokernel} into enclaves. With a LibOS providing system calls, legacy code can run inside enclaves with few or even no modifications.

One highly desirable feature for LibOSes, or any OSes in general, is multitasking. Multitasking is important since virtually any non-trivial application demands more than one process. UNIX has long been known of its philosophy on the rule of composition: design programs to be connected with other programs~\cite{UnixArt}. In the modern era of cloud computing, even a single-purpose, cloud-native application in a container often requires running the main application along with some dependent services (e.g., \texttt{sshd}~\cite{sshd}, \texttt{etcd}~\cite{etcd}, and \texttt{fluentd}~\cite{fluentd}). So multitasking is an indispensable feature.

\begin{figure}[t]\label{figure:occlum}
\newcommand{\myHeight}{4.4cm}

\centering
\begin{subfigure}[t]{0.24\textwidth}
    \centering
    \includegraphics[height=\myHeight]{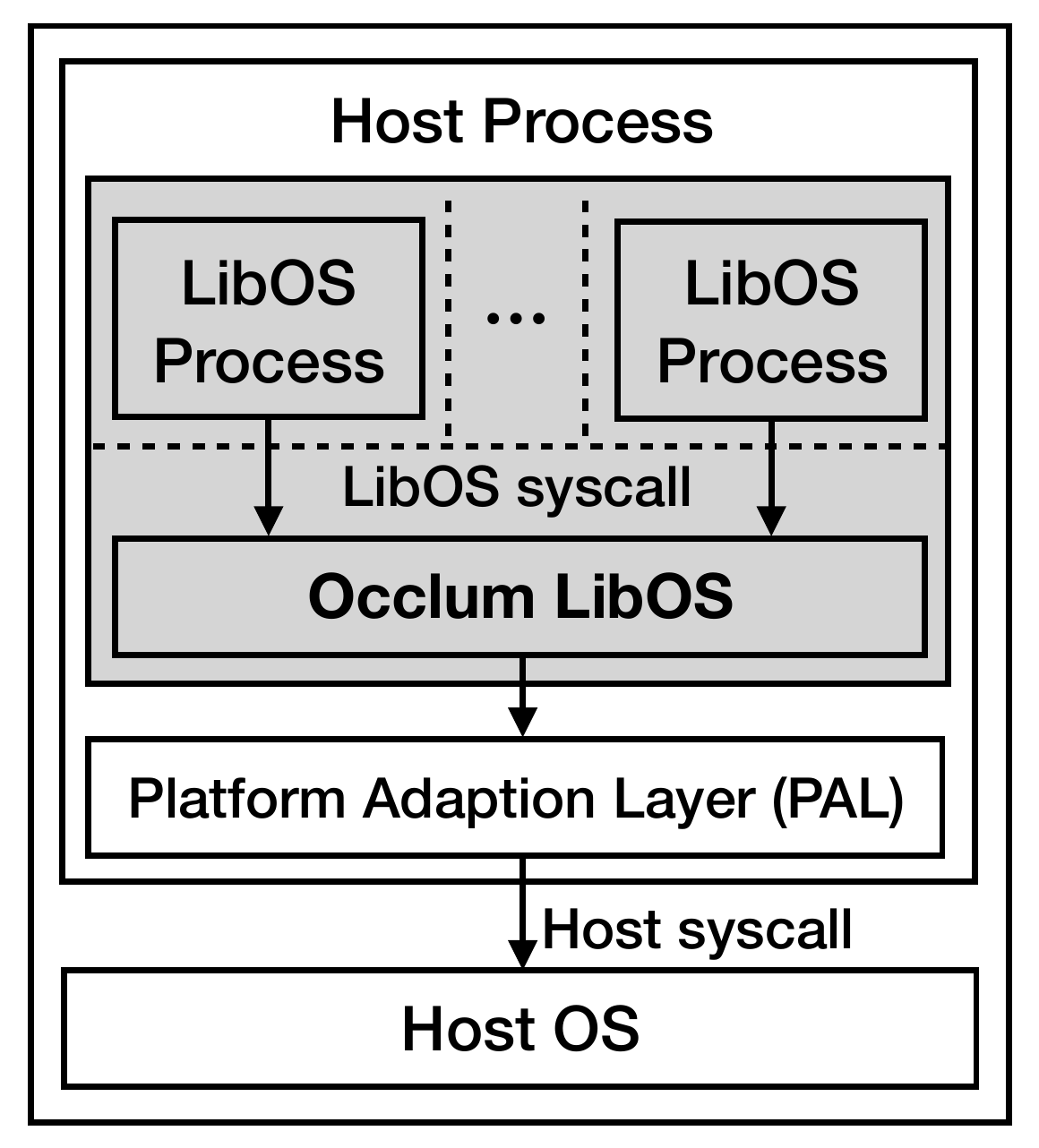}
    \caption{Occlum LibOS}
    \label{figure:occlum:libos}
\end{subfigure}%
~ 
\begin{subfigure}[t]{0.24\textwidth}
    \centering
    \includegraphics[height=\myHeight]{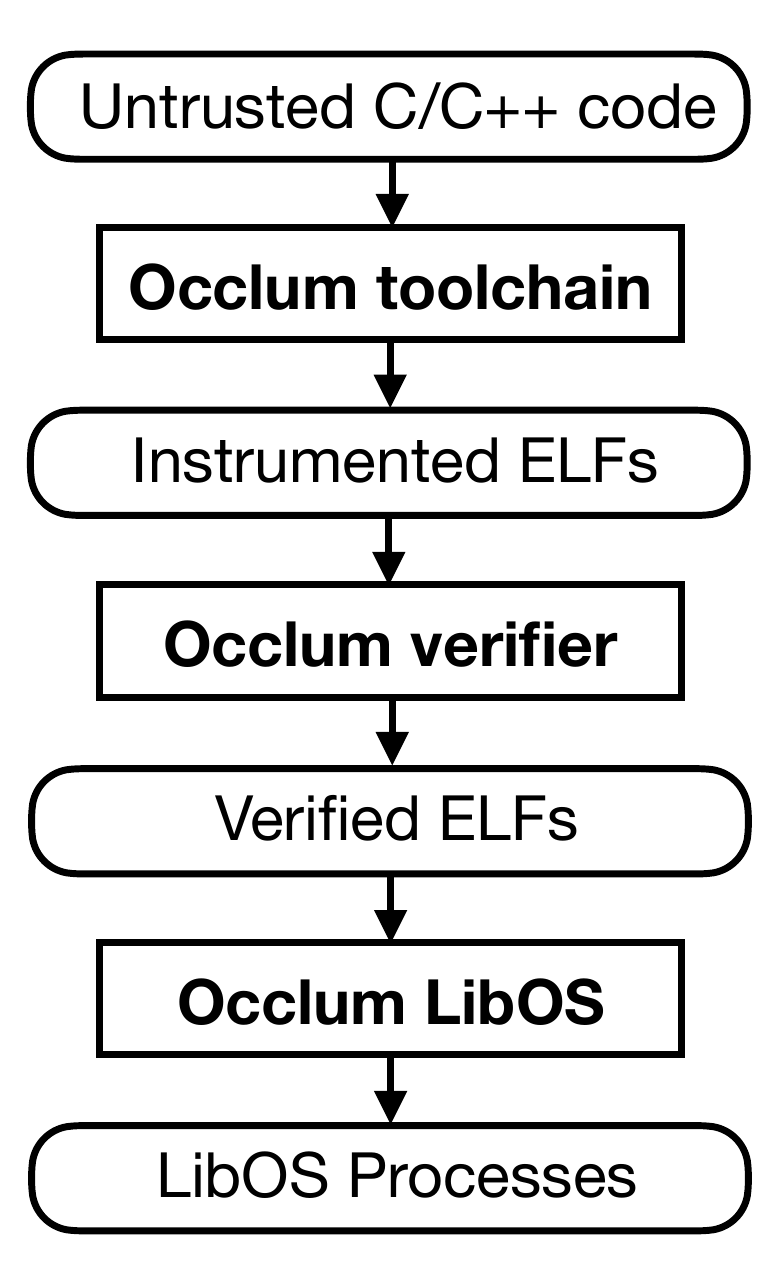}
    \caption{Occlum workflow}
    \label{figure:occlum:workflow}
\end{subfigure}

\caption{An overview of Occlum system, which consists of three components: the toolchain, the verifier, and the LibOS. The gray area represents an enclave.}

\let\myHeight\undefined
\end{figure}

However, existing SGX LibOSes cannot support multitasking both \emph{securely} and \emph{efficiently}. The most advanced, multitasking SGX LibOS is Graphene-SGX~\cite{GrapheneSGX}, which implements LibOS processes as \emph{Enclave-Isolated Processes (EIPs)}. Each EIP is hosted by one instance of the LibOS inside an enclave; that is, $n$ EIPs require $n$ LibOS instances and $n$ enclaves. The strong enclave-based isolation between EIPs, however, causes performance and usability issues.
First, process creation is extremely expensive due to the high cost of enclave creation. Process creation on Graphene-SGX is reported to be nearly $10,000\times$ slower than that on Linux~\cite{GrapheneSGX}.
Second, inter-process communication (IPC) between EIPs is also expensive. EIPs, which are isolated completely by enclave boundaries, have to communicate with each other by transferring encrypted messages through untrusted memory. The encryption and decryption add significant overhead.
Third, synchronizing between multiple LibOS instances is painful. The most notable example is the encrypted file system. As there are multiple LibOS instances, the metadata and data of the file system are thereby spread across multiple enclaves. Thus, maintaining a unified view of the file system across EIPs is difficult and inefficient. This explains why Graphene-SGX lacks a \emph{writable}, encrypted file system.
For the above reasons, it is not clear how to achieve secure and efficient multitasking in SGX LibOSes---up until now.

In this paper, we present Occlum, a system that enables \emph{secure} and \emph{efficient} multitasking in a LibOS for Intel SGX. Different from prior work on SGX LibOSes, we explore an opportunity of synergy between compiler techniques and LibOS design. Specifically, we propose to implement the LibOS processes as \emph{SFI-Isolated Processes (SIPs)}, which reside alongside the LibOS in the single address space of an enclave (see Figure \ref{figure:occlum:libos}). Software Fault Isolation (SFI)~\cite{FirstSFI} is a software instrumentation technique for sandboxing untrusted modules (called \emph{domains}). We design a novel SFI scheme named \emph{MPX-based, Multi-Domain SFI (MMDSFI)}, which, compared with existing SFI schemes, is unique in its support of an \emph{unlimited number} of domains without any constraints on their \emph{addresses} and \emph{sizes}. Thus, we can leverage MMDSFI to implement intra-enclave isolation mechanisms for SIPs, including inter-process isolation and process-LibOS isolation.

To ensure the trustworthiness of Occlum's isolation mechanisms based on MMDSFI, we introduce the Occlum verifier, which is an independent \emph{binary verifier} that takes as input an ELF binary and statically checks whether it is compliant with the security policies of MMDSFI. We describe in depth the design of the verifier and prove its security guarantees mathematically. By introducing the verifier, we exclude Occlum's MMDSFI-enabled toolchain---which is large and complex---from the Trusted Computing Base (TCB) and rely only on the verifier as well as the LibOS for security.

With MMDSFI implemented by the Occlum toolchain and verified by the Occlum verifier, the Occlum LibOS can safely host multiple SIPs inside an single enclave. Since the SIPs reside inside the same address space, there are new opportunities for sharing between SIPs. As a result, the Occlum LibOS can improve both performance and usability of multitasking, achieving fast process startup, low-cost IPC, and writable encrypted file system.

We have implemented the Occlum system, which consists of the three components shown in Figure \ref{figure:occlum:workflow}: the toolchain, the verifier, and the LibOS. Our prototype implementation comprises over $20,000$ lines of source code in total and is available on GitHub~\cite{OcclumGithub}. Experimental evaluation on CPU-intensive benchmarks shows that MMDSFI incurs an average of $36\%$ performance overhead. Despite this overhead incurred by MMDSFI, the Occlum LibOS outperforms the state-of-the-art SGX LibOS on multitasking-heavy workloads by up to $6,600\times$ on micro-benchmarks and up to $500\times$ on application benchmarks. Furthermore, the security benchmark (RIPE~\cite{RIPE}) shows that MMDSFI prevents all memory attacks that may break the isolation of SIPs. These results demonstrate that our SIP-based approach is secure and efficient.

This paper has four major contributions:

1. We propose \emph{SFI-Isolated Processes (SIPs)} to realize secure and efficient multitasking on SGX LibOSes, which is radically different from the traditional approach (\S\ref{section:sips});

2. We design a novel SFI scheme named \emph{MPX-based, Multi-Domain SFI (MMDSFI)} to enforce the isolation of SIPs (\S\ref{section:mmdsfi});

3. We describe in depth our \emph{binary verifier} that statically checks whether an ELF binary is compliant with the security policies enforced by MMDSFI (\S\ref{section:verification}) and present a security analysis against two common classes of attacks (\S\ref{section:security_analysis});

4. We design and implement the \emph{first SIP-based, multitasking LibOS} for Intel SGX (\S\ref{section:libos} and \S\ref{section:implementation}) and demonstrate its performance and security advantages with various benchmarks (\S\ref{section:evaluation}).
\section{Background and Related Work}\label{section:background}

\subsection{Intel Software Guard Extensions (SGX)}

The background knowledge about Intel SGX that is relevant to our discussion is summarized below.

\textbf{Enclave creation.}
During enclave creation, the untrusted OS loads the code and data to the enclave pages, and then the enclave is marked as \emph{initialized}. From this moment, CPU guarantees the protection of the enclave from any code outside the enclave. While an enclave is being created, its contents are cryptographically hashed to calculate the \emph{measurement}. As this process involves a lot of cryptographic computation, it is expensive to create an enclave.

\textbf{Enclave dynamic memory management.}
On SGX 1.0, after an enclave is initialized, enclave pages cannot be added, removed, or modified with their permissions. On SGX 2.0, this restriction has been removed by new SGX instructions. However, Intel has not yet shipped SGX 2.0 CPUs widely. So we implement Occlum on SGX 1.0 to maximize its compatibility.

\textbf{Intra-enclave isolation.}
Currently, there is no hardware isolation mechanisms that are capable or suitable to partition an enclave into smaller security domains. Segmentation is disabled inside enclaves. Page tables are untrusted in SGX's security model. Intel Memory Protection Keys (MPK)~\cite{IntelMPK}) is based on page tables. So MPK is also untrusted to SGX. This is why we turn to SFI for intra-enclave isolation.

\textbf{SGX threads.}
Multiple SGX threads can execute inside an enclave simultaneously. The execution of an SGX thread may be interrupted by hardware exceptions, causing the CPU core of the SGX thread to exit the enclave. This is called an asynchronous enclave exit (AEX). Upon the occurrence of an AEX, the state of the CPU core is automatically stored in a prespecified, secure memory area called state save area (SSA). And later when the SGX thread is about to resume its execution inside the enclave, the state of the CPU core will be stored according to SSA.

\subsection{Library OSes for Intel SGX}

\iffalse
\begin{table}[]
\small
\centering
\caption{A comparison of Occlum with the state-of-the-art LibOSes for SGX.}
\label{table:liboses}
\begin{tabular}{lcccc}
\toprule
	%
	&
	Haven
	&
	Scone
	&
	Graphene
	&
	Occlum
	\\
\midrule
\textbf{Legacy binaries}
	&
	Yes
	&
	No
	&
	Yes
	&
	\textbf{No}
	\\
\textbf{Small TCB}
	&
	No
	&
	Yes
	&
	Yes
	&
	\textbf{Yes}
	\\
\textbf{Memory safety}
	&
	No
	&
	No
	&
	No
	&
	\textbf{Yes}
	\\
\iffalse
\textbf{Sandboxing}
	&
	No
	&
	No
	&
	No
	&
	\textbf{Yes}
	\\
\fi
\textbf{Multitasking}
	&
	No
	&
	No
	&
	Inefficient
	&
	\textbf{Efficient}
	\\
\bottomrule
\end{tabular}
\end{table}
\fi

We compare Occlum with the state-of-the-art LibOSes for SGX---i.e., Haven~\cite{Haven}, Scone~\cite{Scone}, Panoply~\cite{Panoply}, and Graphene-SGX~\cite{GrapheneSGX}---in three dimensions.
%The result is summarized in table \ref{table:liboses}.

\textbf{Compatibility.}
An SGX LibOS may be compatible with legacy applications at either binary or code level. 
Haven and Graphene-SGX are binary level compatible while Scone, Panoply and Occlum are source code level compatible via cross compilers.

%-compatible LibOSes, e.g., Haven and Graphene-SGX, support legacy binaries.
%Code-compatible LibOSes, e.g., Scone, Panoply and Occlum, support legacy source code via a cross compiler.

We argue that \emph{the code-level compatibility is acceptable for most use cases} since it has eliminated most of the efforts required to port applications for SGX.
And even a binary-compatible SGX LibOS sometimes demands recompilation so that the legacy source code can be modified to work around the hardware limitatations of SGX or the system call limitations of the LibOS.
Futhermore, the recompilation of source code provides the opportunity to integrate SGX-specific, compiler-based hardening techniques~\cite{Varys, SGXBounds}.
	% Varys: Protecting SGX enclaves from practical side-channel attacks

\textbf{Multitasking.}
Existing SGX LibOSes cannot support multitasking both securely and efficiently. Haven supports multitasking in a single-address-space architecture like Occlum, but it lacks isolation for its LibOS processes. Scone, Graphene-SGX, and Panoply support multitasking with EIPs, but suffer from performance and usability issues (\S\ref{section:sips:advantages}).

\textbf{TCB sizes.}
As Haven reuses the source code of Drawbridge~\cite{Drawbridge}, it ends up with a huge TCB. The other four LibOSes including Occlum are written from scratch and thus absent of unnecessary OS functionalities. This results in small TCBs and thus small attack surfaces.

\subsection{Intel MPX and SFI}
Intel Memory Protection Extensions (MPX)~\cite{IntelMPX} is a set of extensions to the x86 instruction set architecture to provide bound checking at runtime. It provides four bound registers \Bnd{0} - \Bnd{3}. Each bound register stores a pair of 64-bit values, one for the lower bound and the other for the upper bound. Two instructions, \texttt{bndcl} and \texttt{bndcu} are introduced to check a given address against a bound register's lower and upper bound, respectively. If the check fails, an exception will be raised by the CPU. These four bound registers are saved when AEX occurs and are restored when the enclave resumes its execution from AEX~\cite{IntelSGX}. Occlum does not use MPX's expensive bound tables or bound table-related instructions.

Software Fault Isolation (SFI)~\cite{FirstSFI} is a software instrumentation technique for sandboxing untrusted modules (called domains). Existing SFI schemes have limitations on the number, addresses, or sizes of domains to simplify their designs and minimize the overheads. For example, the well-studied SFI scheme Native Client~\cite{NaCl-x64} requires two 40GB unmapped memory regions around a 4GB-size domain. PittSFIeld~\cite{PittSFIeld} only supports at most $64 - n$ domains of $n$-bit address space on x86-64.

Our SFI scheme MMDSFI intends to sandbox a (potentially) large number of processes inside the limited address space of an enclave. MMDSFI aims to put no constraints on the number, addresses, or sizes of domains. We find two advantages of using MPX registers for domain bounds:

1. Bound registers can represent any address or size for a domain.

2. During thread switching, bound registers are automatically saved and stored by CPU. Thus, the maximum number of domains does not depend on the number of bound registers, but the address size of an enclave.

Existing MPX-enabled systems~\cite{NoNeedToHide,Apparition} use MPX simply for reducing the overhead of SFI. MMDSFI fully leverages MPX's advantages to achieve the flexibility on the number, addresses, and sizes of domains.
\section{SFI-Isolated Processes (SIPs)}\label{section:sips}

\begin{table}[]
\small
\centering
\caption{A comparison between SIPs and EIPs.}
\label{table:eips_vs_sips}
\begin{tabular}{p{2.9cm}cc}
\toprule
	&
	\begin{minipage}{2.2cm}\centering
    	\textbf{EIPs}
    	\vspace*{0.1cm}
    \end{minipage}
	&
	\begin{minipage}{2.2cm}\centering
    	\textbf{SIPs}
    \end{minipage}
	\\
	&
    \begin{minipage}{2.0cm}\centering
        \includegraphics[scale=0.32]{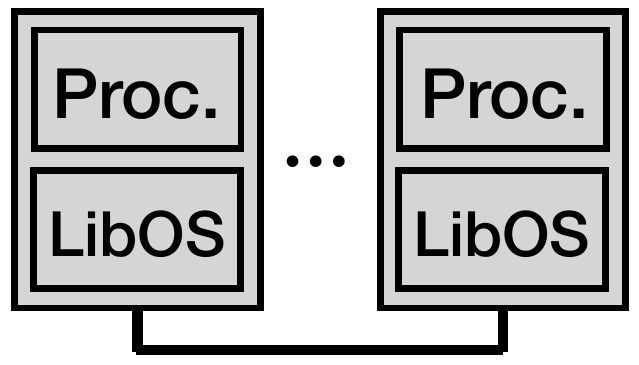}
    \end{minipage}
    &
    \begin{minipage}{2.0cm}\centering
        \includegraphics[scale=0.32]{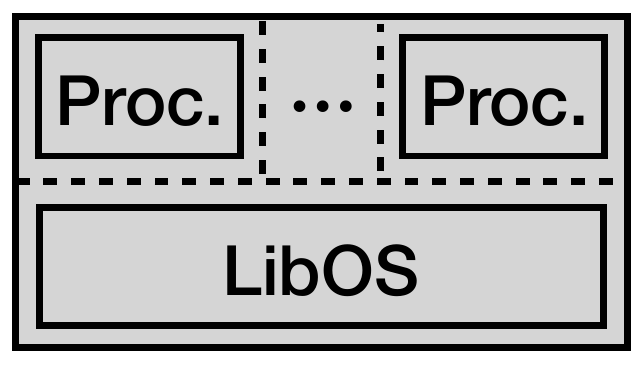}
    \end{minipage}
    \\
\midrule
\textbf{Process creation}
    &
    Expensive
    &
    \textbf{Cheap}
    \\
\textbf{IPC}
    &
    Expensive
    &
    \textbf{Cheap}
    \\
\textbf{Shared file systems}
    &
    Read-only
    &
    \textbf{Writable}
    \\
\bottomrule
\end{tabular}
\end{table}

In this section, we give an overview of SIPs by first describing its threat model, then highlighting its advantages over traditional EIPs, and finally discussing its feasibility on SGX.

\subsection{Threat Model and Security Goals}\label{section:sips:threat_model}
We assume that attackers can take full control over the hypervisor, host OS, and host applications on the target machine. SIPs are assumed to be benign (at least when loaded by the LibOS), otherwise their binaries would have been rejected by our verifier. 
However, SIPs may have security vulnerabilities and get compromised by attackers.
In short, we assume a powerful attacker who controls both the infrastructure on which an enclave is running and some malicious SIPs inside the enclave.

To defend against such a powerful attacker, Occlum aims to isolate in-enclave SIPs. Specifically, we enforce the following two kinds of isolation: 1) \emph{Inter-process isolation}, which protects an SIP from other SIPs; and 2) \emph{Process-LibOS isolation}, which protects the LibOS itself from any SIP.

We assume our LibOS is implemented correctly, which is the (runtime) TCB of Occlum. Iago attacks~\cite{Iago} are not considered, which can be addressed by orthogonal work like Sego~\cite{Sego}. Like other SGX LibOSes, Occlum does not hide file access patterns. We do not consider denial-of-service attacks.

We do not consider side-channel attacks in this paper. Side-channel attacks have been shown to be a real threat, especially to Intel SGX~\cite{SGXMemorySideChannel, Foreshadow, SgxPectre}. This field is moving fast: new attacks have been kept being proposed, so have new defenses~\cite{Varys, TSGX}. We believe eventually these efforts that are independent from ours will be able to provide adequent defense against side-channel attacks. Also, covert-channel attacks are out of the scope of this paper.

\subsection{Advantages of SIP}\label{section:sips:advantages}

SIPs have performance and usability advantages compared to the traditional EIPs. A comparison between SIPs and EIPs is summarized in Table \ref{table:eips_vs_sips} and explained below.

\textbf{Process creation.}
Creating a new EIP requires three steps: (1) creating a new enclave, (2) doing local attestation with other enclaves, and (3) duplicating the process state over an encrypted stream. In contrast, creating an SIP does not involve any of the three steps. Thus, SIPs are significantly cheaper to create than EIPs.

\textbf{IPC.}
IPCs between EIPs are typically implemented by sending and receiving encrypted data through untrusted buffers outside the enclave. Yet, IPC between SIPs is simply copying data from one SIP to another, with no encryption involved.

\textbf{Shared file system.}
In the traditional EIP-based approach, there are multiple instances of the LibOSes communicating via secure communication channels. This setup must face the challenge of data synchronization. To avoid this difficulty, traditional LibOSes like Graphene-SGX only support a read-only, encrypted file system. In contrast, all SIPs inside an enclave share the same instance of the LibOS. Thus, a writable, encrypted file system can be implemented relatively easily.

\subsection{Spawn Instead of Fork}\label{section:sips:feasibility}

\texttt{fork} system call requires a forked child process to have the same address space as its parent process. This semantic is incompatible with single-address-space OSes~\cite{SASOS}, including Occlum, where all processes reside in a single address space. So, in Occlum, processes are created with the \texttt{spawn} system call, which is similar to libc's \texttt{posix\_spawn} and Solaris's \texttt{spawn}~\cite{SolarisSpawn}.

Most uses of \texttt{fork} and \texttt{exec} can be readily replaced with \texttt{spawn}. As for those applications such as Apache, PostgreSQL, and Redis that use \texttt{fork} alone, the existence of their Windows ports or versions~\cite{ApacheWindows, PostgreSQLWindows, RedisWindows} implies that \texttt{fork}-based code can be rewritten to use \texttt{spawn}-like APIs. Our experience in porting multi-process applications for Occlum also confirms that replacing \texttt{fork} with \texttt{spawn} is not as hard as one might expect (\S\ref{section:evaluation:application_benchmarks}).

Putting the compatibility issues aside, \texttt{spawn} is considered superior to \texttt{fork}. A recent paper~\cite{AForkInTheRoad} reflects on the pros and cons of \texttt{fork} and concludes that ``we should acknowledge that \texttt{fork}'s continued existence as a first-class OS primitive holds back systems research, and deprecate it.
\section{MPX-Based, Multi-Domain SFI (MMDSFI)}\label{section:mmdsfi}

\begin{figure*}[t!]\label{figure:mmdsfi}
\newcommand{\myHeight}{3.8cm}

\centering
\begin{subfigure}[t]{0.225\textwidth}
    \centering
    \includegraphics[height=\myHeight]{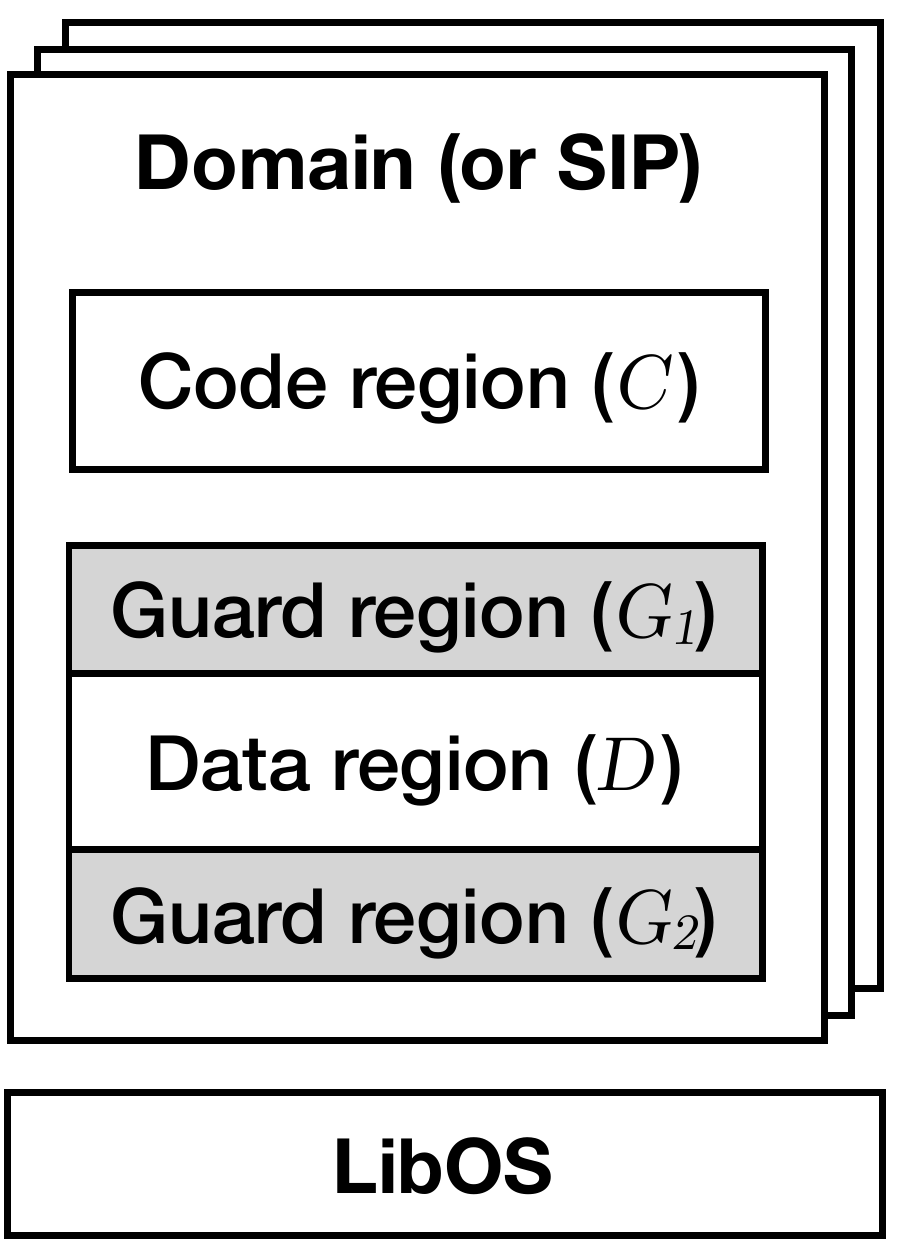}
    \caption{The memory layout}
    \label{figure:mmdsfi:memory_layout}
\end{subfigure}%
~ 
\begin{subfigure}[t]{0.390\textwidth}
    \centering
    \includegraphics[height=\myHeight]{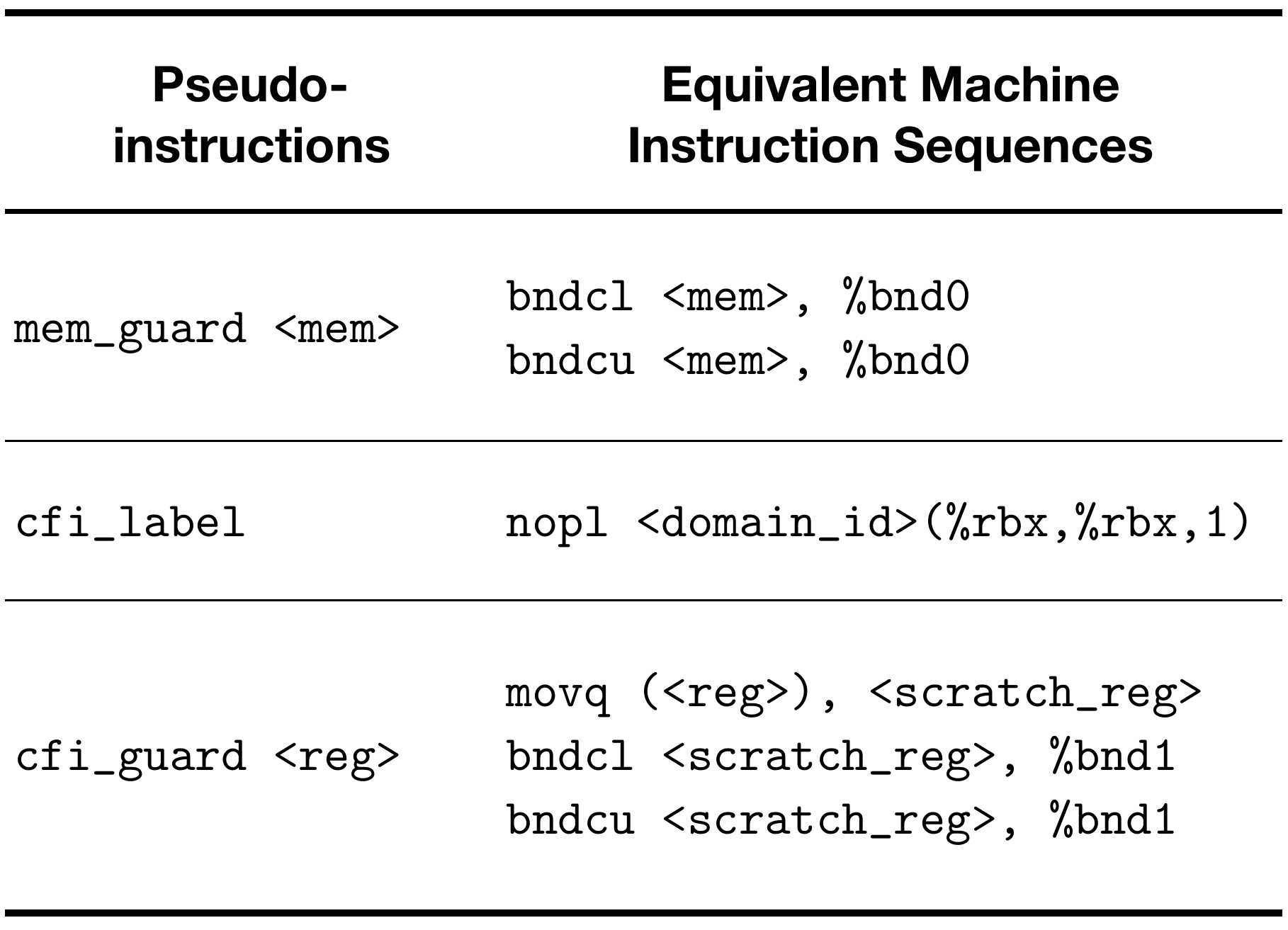}
    \caption{The pseudo-instructions}
    \label{figure:mmdsfi:pseudo-instructions}
\end{subfigure}
~ 
\begin{subfigure}[t]{0.350\textwidth}
    \centering
    \includegraphics[height=\myHeight]{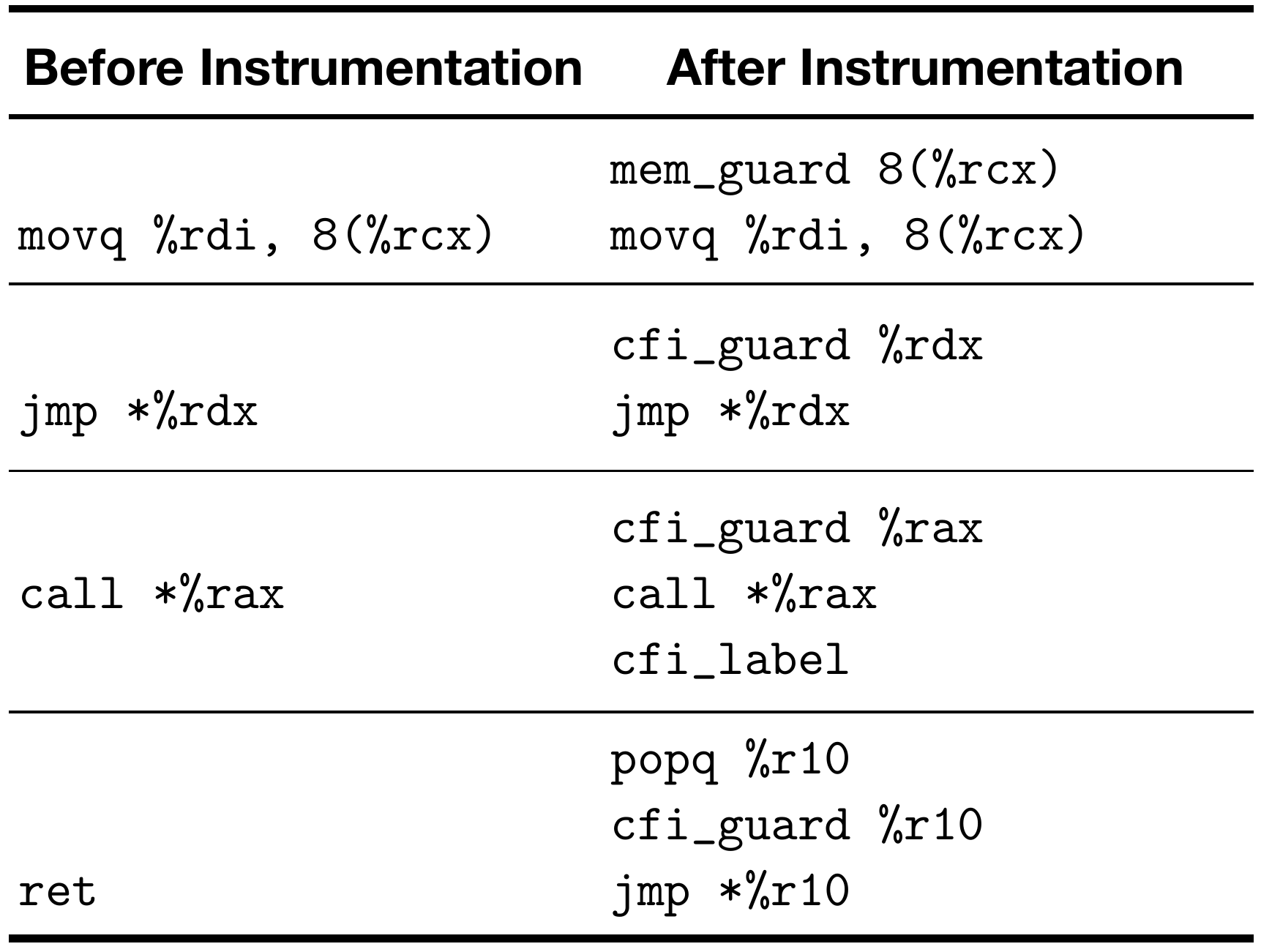}
    \caption{Instrumentation examples}
    \label{figure:mmdsfi:examples}
\end{subfigure}

\caption{The design of MMDSFI. All assembly code follows the AT\&T syntax. The bound register \Bnd{0} and \Bnd{1} are initialized by the LibOS to the range $[\Dbegin, \Dend)$ and $[$\CfiLabel, \CfiLabel\!\!$]$, respectively. The operand \RegOperand denotes a general-purpose register; \ScratchReg, a scratch register for storing intermediate values; \MemOperand, a memory address; \DomainId, the ID of the domain that the code belongs to.} 

\let\myHeight\undefined
\end{figure*}

%In this section, we present a novel SFI scheme named MPX-based, Multi-Domain SFI (MMDSFI), with which Occlum LibOS can enforce the isolation of SIPs.

Occlum uses a novel SFI scheme named MPX-based, Multi-Domain SFI (MMDSFI) to enforce the isolation of SIPs.

\subsection{Overview}\label{section:mmdsfi:overview}

MMDSFI requires a memory layout shown in Figure \ref{figure:mmdsfi:memory_layout}. A domain of MMDSFI has two main memory regions: the region $C$ for the user code, and the region $D$ for the user data. Region $C$ and $D$ are mapped to enclave pages with \texttt{RWX} and \texttt{RW} permissions, respectively. Region $D$ is surrounded by two guard regions $G_1$ and $G_2$, which are of the same size and not mapped to any enclave pages, thus triggering exceptions when accessed. These guard regions, as a common technique of SFI, are introduced to simplify the instrumentation (\S\ref{section:mmdsfi:instrumentation}) and facilitate some optimizations (\S\ref{section:mmdsfi:optimizations}).
Multiple domains can coexist in a single address space, and the memory ranges of these domains are exclusive to each other. All domains share a single instance of the LibOS, which is in charge of managing the domains and processing system calls from them.
% TODO: explain why guard regions facilitates instrumentation

As an SFI scheme, the security goal of MMDSFI is to \emph{sandbox untrusted user code}. The untrusted code is first compiled---with the instrumentation required by MMDSFI---into a binary. Then, the binary is loaded by the LibOS into a domain: the code of the binary is loaded into the region $C$ and the data into the region $D$. This domain is then run as an SIP (so actually \emph{the two terms of domain and SIP can be used interchangeably for our discussion}). Together with the compile-time instrumentation of MMDSFI and the runtime support of the LibOS, we can guarantee that the untrusted user code is sandboxed. Specifically, the following two security policies are enforced:

\textbf{The memory access policy.} For any memory access instruction $I$ in $C$, $I$ must access the memory within range $[\Dbegin, \Dend)$, where $\Dbegin$ and $\Dend$ are addresses where the region $D$ begins and ends.

\textbf{The control transfer policy.} For any control transfer instruction $I$ in $C$, $I$ must target an address within $[\Cbegin, \Cend)$, where $\Cbegin$ and $\Cend$ are the addresses where the region $C$ begins and ends.

If an instruction violates the security policies above, then the instruction is \emph{invalid}. Our goal is to guarantee that any \emph{invalid} instruction is either \emph{prevented} or \emph{detected}.

\iffalse
Before describing more details about MMDSFI, we want to make clear the assumptions that our discussion in the rest of this section rely on. To simplify the presentation, we assume the untrusted code runs inside enclaves and in the 64-bit mode of an x86-64 CPU, although our SFI scheme can be easily generalized for use outside enclaves or in the 32-bit mode. Furthermore, we assume the LibOS is trusted and can provide all the runtime support required by MMDSFI, e.g., initializing the memory of a domain correctly, processing system calls safely (so that system calls cannot be abused by the untrusted code inside domains to compromise the security guarantee of MMDSFI), and handling the exceptions generated from MMDSFI properly.
\fi
\subsection{Instrumentation}\label{section:mmdsfi:instrumentation}

MMDSFI instruments the untrusted code by inserting checks for or rewriting unsafe instructions to enforce the aforementioned security policies.

\textbf{Confining memory accesses.} To this end, we introduce \MemGuard pseudo-instruction, which takes a memory operand \MemOperand and checks whether \MemOperand is within the data range $[\Dbegin, \Dend)$ of the current domain. If the check passes, it does nothing; otherwise, it triggers an exception, which can then be captured by the LibOS. As shown in Figure \ref{figure:mmdsfi:pseudo-instructions}, \MemGuard pseudo-instruction can be implemented efficiently with two MPX bound check instructions. As the first example shown in Figure \ref{figure:mmdsfi:examples}, MMDSFI inserts \MemGuard before every unsafe memory access instruction so that its memory accesses are confined within $[\Dbegin, \Dend)$.

\textbf{Confining control transfers.} Before describing how MMDSFI confines control transfers, let's take a few relevant observations. First, since x86 instructions are variable-length, a faulty control transfer could jump into the middle of (pseudo-)instructions and execute unpredictable instructions. This would completely jeopardize the validity of SFI and thus must not be allowed. Second, some instruction sequences must be treated as a whole. One such example is \MemGuard and its guarded memory access instruction, which must be executed as a whole. Otherwise, jumping in between the two instructions would skip the \MemGuard, thereby bypassing the memory access confinement. So, this, too, must not be allowed. Third, \emph{indirect} control transfer instructions need runtime checks while \emph{direct} control transfer instructions do not. A direct control transfer instruction (e.g., \texttt{jmp \ImmdOperand}, where \ImmdOperand denotes an immediate value) has its target address hard-coded in the instruction and thus can be verified at compile time. Yet, an indirect control transfer instruction (e.g., \texttt{jmp \RegOperand}) gives its target address in a register or a memory location, which cannot be determined until runtime. Thus, indirect control transfer instructions need runtime checks.

Based on the above observations, we enforce a coarse-grained control-flow integrity (CFI) ~\cite{FirstSFI} in MMDSFI by introducing a pair of pseudo-instructions: \CfiLabel and \CfiGuard, as shown in Figure \ref{figure:mmdsfi:pseudo-instructions}.

\CfiLabel has the following interesting properties:

(1) No operation. It has no visible impact on CPU;

(2) Alignment. The encoding has a fixed length of 8 bytes;

(3) Nonexistence. The first 4 bytes of the encoding does not appear anywhere in the uninstrumented code, including both the untrusted user code and the trusted LibOS code;

(4) Uniqueness. The last 4 bytes of the encoding is the unique ID of the domain where the \CfiLabel resides.

To satisfy these properties, we implement \CfiLabel with a special 8-byte \texttt{nop} instruction, as shown in Figure \ref{figure:mmdsfi:pseudo-instructions}. This \texttt{nop} is not in its most common form, so it is not supposed to be emitted by any off-the-shelf compiler. The \texttt{nop} is encoded in 8 bytes with the last 4 bytes being any 32-bit value of our choice. So, when loading instrumented code into a domain, the LibOS rewrites the last 4 bytes of all \CfiLabels to the value of the ID of the current domain. This is to satisfy the uniqueness property required above.

With the properties above, we can use \CfiLabels to ``label'' the valid targets of indirect control transfers. Specifically, MMDSFI inserts a \CfiLabel at every valid target address of indirect control transfers (e.g., the start address of a function). And only at these addresses can \CfiLabels be found. Thus, the inserted \CfiLabels can be used at runtime to verify indirect control transfers by the pseudo-instruction that is to be introduced below. 

\CfiGuard pseudo-instruction checks whether the address given in a register operand \RegOperand is a valid target of indirect control transfers. If it is a valid target, \CfiGuard does nothing; otherwise, it raises an exception. The validity of a target address can be easily determined with \CfiLabels. As shown in Figure \ref{figure:mmdsfi:pseudo-instructions}, \CfiGuard is implemented in a sequence of three instructions: the first \texttt{mov} loads the value at the target address into a scratch register \ScratchReg, then two following bound checks compare the value in \ScratchReg with \Bnd{1}, which has been set by the LibOS to the range $[$\CfiLabel, \CfiLabel\!\!$]$. So the two bound checks are essentially a test for equality. If and only if the value at the target address equals to \CfiLabel can the target be valid. With \CfiGuards, we now can confine indirect control transfers to target only valid addresses. See figure \ref{figure:mmdsfi:examples} for some typical examples of instrumentation.

Using the three pseudo-instructions, we can instrument any unsafe memory access or control transfer instruction as shown in figure \ref{figure:mmdsfi:examples}, thus enforcing the security policies of MMDSFI. The correctness of MMDSFI is discussed formally in \S\ref{section:verification}.

\subsection{Optimizations}\label{section:mmdsfi:optimizations}

A naive implementation of MMDSFI instrumentation has to insert a \MemGuard for every memory accesses. This can easily lead to an intolerable performance penalty. However, thanks to our enforcement of CFI, we can employ standard dataflow analysis~\cite{DragonBook} on the control-flow graph (CFG) to seize optimization opportunities enabled by the guard regions (\S\ref{section:mmdsfi:overview}). Specifically, we can use \emph{range analysis} to track the ranges of possible values in registers before and after each instruction, and perform the following two optimizations~\cite{SFIwithCFI}:

(1) \emph{Redundant check elimination}. Recall that the data region $D$ is surrounded by the guard regions $G_1$ and $G_2$ that will trigger exceptions if accessed. So, if a memory address $x$ has been confined within the data range $[\Dbegin, \Dend)$, then any memory address determined by the range analysis to be within $[x - G_i.size, x + G_i.size]$ is also safe and requires no check via \MemGuard, where $G_i.size$ is the size of a guard region.

(2) \emph{Loop check hoisting}. In a loop, if there is a \MemGuard whose memory operand \MemOperand is increased or decreased by a constant value smaller than $G_i.size$ per loop iteration, then a new \MemGuard can be inserted before the loop and the original, in-loop \MemGuard can be eliminated due to the same reason in the first optimization. The net result is that the \MemGuard will be executed only once per loop instead of once per loop iteration.

While more optimizations based on range analysis are possible, we found these two optimizations are sufficient to reduce the overhead to an acceptable level (\S\ref{section:evaluation:mmdsfi_benchmarks}).
\section{Binary Verification}\label{section:verification}

\begin{algorithm}[h]
\small
\footnotesize
\newcommand{\addr}{\mathit{addr}}
\newcommand{\instr}{\mathit{instr}}
\SetArgSty{}
\KwIn{$C$, the code segment of an ELF binary}
%  by byte scan}
\KwOut{$R$, the set of all reachable instructions in $C$}
$R \leftarrow \{\}$\;
$S \leftarrow$ Get all the addresses of \texttt{cfi\_label}s by scanning $C$ byte by byte\;
\While{$S \neq \{\}$}{
	$\addr \leftarrow $ Pop an item from $S$\;
	\While{true} {
		\If{$\addr$ \text{is not within} $C$} {
			abort\;
		}
		\BlankLine

		$\instr \leftarrow$ Disassemble the instruction at $\addr$\;
		\uIf{$\instr$ is invalid}{
			abort\;
		}
		\uElseIf{$\instr \in R$}{
			break\;
		}
		\ElseIf{$\instr$ overlaps with any $i \in R$}{
			abort\;
		}

		$R \leftarrow R \cup \{ \instr \}$\;
		\BlankLine

		\If{$\instr$ is a direct control transfer}{
			$\mathit{target\!\_addr} \leftarrow$ Calculate the target of $\instr$\;
			$S \leftarrow S \cup \{ \mathit{target\!\_addr} \}$\;
		}
		\If{$\instr$ is an unconditional control transfer}{
			break\;
		}
		$\addr \leftarrow \addr + \instr.\mathit{length}$\;
	}
}
% add new keywords
\caption{Stage 1 disassembles the binary completely and reliably, generating $R$ the set of all reachable instructions.}\label{algorithm:disassembly}
\label{algorithm:verification:stage1}
\end{algorithm}

\newcommand{\myWidth}{0.46\textwidth}

\begin{figure}[t!]
    \centering
    \includegraphics[width=\myWidth]{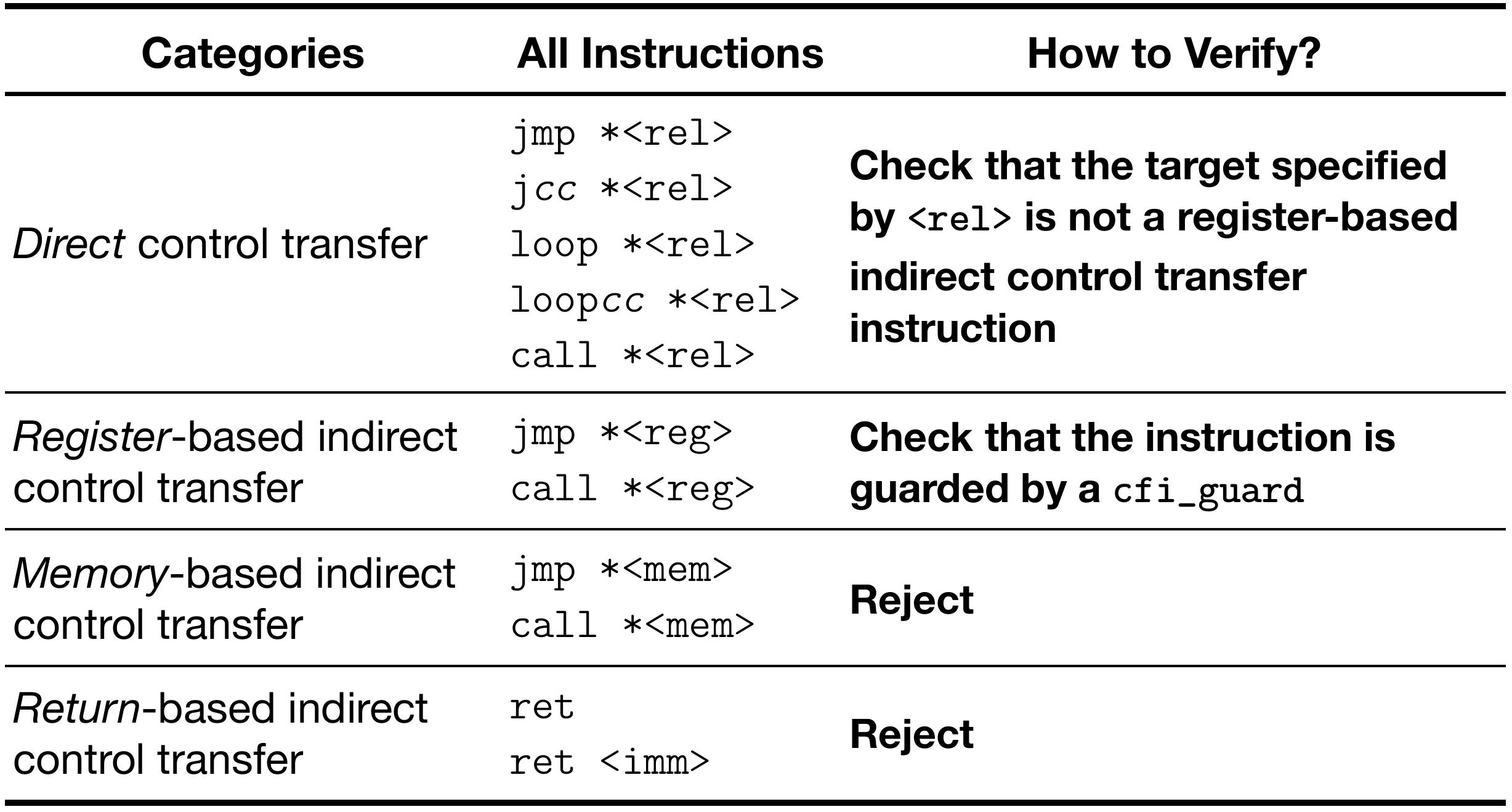}
    \caption{Stage 3 verifies control transfer instructions by classifying them into four categories.}
    \label{figure:verification:stage3}
\end{figure}
\begin{figure}[t!]
    \centering
    \includegraphics[width=\myWidth]{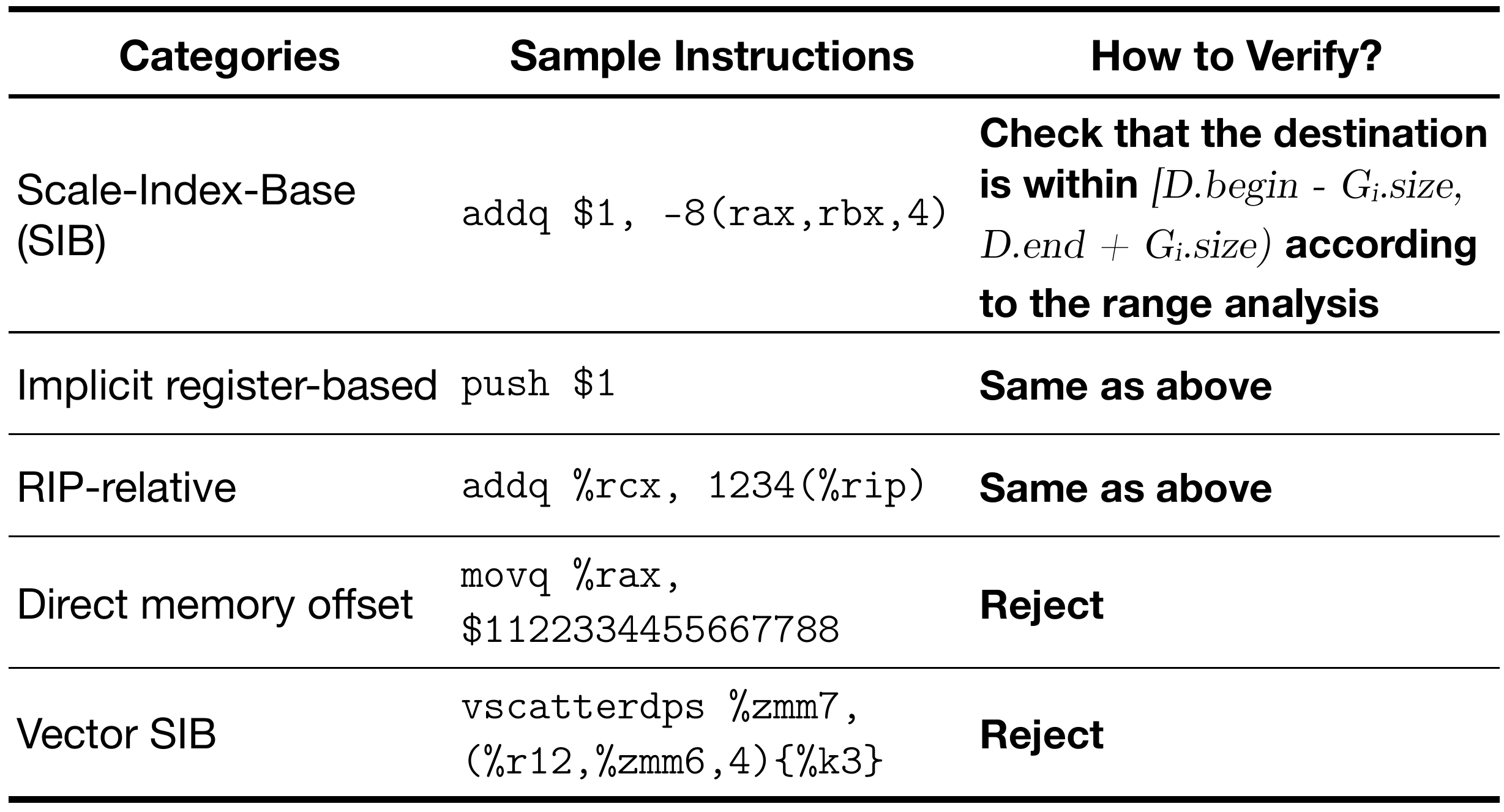}
    \caption{Stage 4 verifies memory access instructions by classifying them into five categories. Direct memory offset, which can hard-code a 64-bit memory address in \texttt{mov}, is rejected because no fixed addresses can be assumed to be within a domain. Vector SIB is rejected because it allows one instruction to access to multiple non-contiguous memory locations. }
    \label{figure:verification:stage4}
\end{figure}
\let\myWidth\undefined
\let\myHeight\undefined

To ensure the trustworthiness of Occlum's security properties based on MMDSFI, we introduce the Occlum verifier, which is an independent binary verifier that takes an ELF binary as input and statically checks whether it is compliant with the security policies of MMDSFI.

By introducing the Occlum verifier, we exclude the Occlum toolchain from the TCB. This is desirable for two reasons. First, the implementation of an SFI, including MMDSFI, is error-prone since it involves dealing with the low-level details of machine code (e.g., x86 has approximately 1000 distinct instructions) and it is usually implemented upon the huge codebase of a compiler (e.g., GCC and LLVM have millions of lines of source code). Second, even if the Occlum toolchain can be proven flawless, the Occlum verifier can still be very useful. In some interesting use cases~\cite{InfoReleaseConfine, Ryoan}, the %binaries to be loaded by Occlum LibOS are 
user may want to use binaries from untrusted origins. So these binaries can be generated by some arbitrary toolchains; and the Occlum toolchain being flawless is irrelevant. Thus, we must verify the binaries before loading them into enclaves.

The verifier consists of four stages: (1) complete disassembly, (2) instruction set verification, (3) control transfer verification, and (4) memory access verification. If and only if all the four stages pass, can the input binary be claimed to be compliant with MMDSFI and safe to run upon the Occlum LibOS inside an enclave. 

\textbf{Stage 1 - Complete disassembly.} This stage aims to find out every instruction \emph{reachable}, i.e., any instruction that may be executed as part of the untrusted user program. To this end, it disassembles the input ELF binary and generates the set of all reachable instructions, which is denoted as $R$. Our x86 disassembler is extended with the three pseudo-instructions introduced in the last section. So an instruction in $R$ can be either a real x86 instruction or a pseudo-instruction, which is not distinguished for our discussion. Every instruction in $R$ is uniquely identified by its memory address. This resulting set $R$ is the subject of the verification to be carried out in the remaining four stages.

Disassembling an \emph{arbitrary} binary \emph{statically}, \emph{reliably}, and \emph{completely} is impossible; existing disassembly algorithms are based on heuristics, providing only best-effort results~\cite{DisassemblerA, DisassemblerB,Depth}. However, thanks to the introduction of \CfiLabel, our disassembly algorithm can give the \emph{completely accurate} results as long as the input binary can eventually pass all the remaining stages of the verifier. (And for those binaries that are to be rejected by the verifier, it does not really matter if the disassembly results are accurate or not.)

The disassembly algorithm is shown in figure \ref{algorithm:verification:stage1}. Firstly, the disassembler gets all \CfiLabels by scanning the code section of the binary byte by byte (line 2). Guaranteed by the LibOS, the user program must start its execution from \CfiLabels. From each of these starting points (line 4), the disassembly algorithm follows the sequential execution of the program (line 21) and every direct control transfer (line 17). Note that the \emph{indirect} control transfers can only target at \CfiLabels (due to the control transfer policy to be verified in Stage 3), which are the starting points themselves. So the branches of indirect control transfers have been implicitly covered. As we have started from every possible starting point and followed every possible branch, we end up with $R$ containing all reachable instructions.

\textbf{Stage 2 - Instruction set verification.} The goal of this stage is to make sure that $R$ does not include any dangerous instructions that can perform privileged tasks that are meant for the LibOS. The dangerous instructions can be classified into the following three categories:

(1) SGX instructions, e.g., \texttt{eexit}, which exits the current enclave, and \texttt{emodpe}/\texttt{eaccept}, which extends/restricts the memory permissions of enclave pages at runtime;

(2) MPX instructions, e.g., \texttt{bndmk} and \texttt{bndmov}, which modify the values of MPX bound registers;

(3) Miscellaneous instructions, e.g., \texttt{xrstor}, which can enable or disable some CPU features including MPX, and \texttt{wrfsbase}/\texttt{wrgsbase}, which modifies \texttt{FS}/\texttt{GS} segment bases.

Note that these dangerous instructions are not supposed to be used in normal user programs. So forbidding them does not restrict the functionality of user programs. The implementation of this stage is quite straightforward: simply scan $R$ to check that no such dangerous instructions exist.

\textbf{Stage 3 - Control transfer verification}. This stage verifies every control transfer instructions in $R$ by classifying it into one of four categories and checking it with the criteria of its belonging category as shown in Figure \ref{figure:verification:stage3}. 

If a binary passes Stage 1-3 of the verification and is loaded into a domain by the LibOS, then we can prove the lemma and theorem below.

\begin{lemma}\label{lemma:control_transfers}
Any register-based indirect control transfer in the domain can only jump to the \CfiLabels in the domain.
\end{lemma}

\begin{prf}
To prove it by contradiction, assume there exists an occurrence of register-based indirect control transfer that violates the lemma.

Let $S = \{I_1, \cdots, I_n\}$ be the CPU execution trace of the domain up until the violation occurs, where $I_1$ is a \CfiLabel, $I_n$ is the violating register-based indirect control transfer, and $n \geq 2$. Without loss of generality, we can assume that $I_n$ is the first violation in the trace; that is, no $1 < j < n$ such that $I_j$ is also an occurrence of register-based indirect control transfer that violates the lemma. Otherwise, we can simply restart our analysis on $I_j$ instead of $I_n$.

Our verification in Stage 3 has checked that $I_n$ is properly guarded with a \CfiGuard, yet still $I_n$ jumps to a location other than some \CfiLabel in the domain. This implies that $I_{n-1}$ cannot be the \CfiGuard and must be a control transfer instruction that jumps to $I_n$. Yet, such $I_{n-1}$ is impossible according to the exhaustive case analysis below:

	(1) $I_{n-1}$ cannot be a \emph{direct} control transfer since our verification in Stage 3 has checked that any direct control transfer does not jump to a register-based indirect control transfer instruction;

	(2) $I_{n-1}$ cannot be a \emph{register-based} indirect control transfer otherwise it would be a violation that happened before $I_n$, which contradicts our assumption that $I_n$ is the first;

	(3) $I_{n-1}$ cannot be a \emph{memory-based} indirect control transfer as our verification in Stage 3 rejects it;

	(4) $I_{n-1}$ cannot be a \emph{return-based} indirect control transfer as our verification in Stage 3 rejects it.

By contradiction, we prove the lemma.
\end{prf}

\begin{theorem}\label{theorem:control_transfers}
Any control transfer in the domain is compliant with the control transfer policy of MMDSFI (\S\ref{section:mmdsfi:overview}).
\end{theorem}

\begin{prf}
For any control transfer instruction in the domain, it must belong to one of the four categories listed in Figure \ref{figure:verification:stage3}. The first category is covered implicitly by the complete disassembly in Stage 1; the second category is implied by Lemma \ref{lemma:control_transfers}; the third and the fourth categories are forbidden by the verification in Stage 3. Thus, the theorem stands in all cases.
\end{prf} 

\textbf{Stage 4 - Memory access verification.} With the CFI of $R$ verified in Stage 3, this stage first builds the CFG for $R$, then leverages this CFG to do the \CfiLabel-aware range analysis described in \S\ref{section:mmdsfi:optimizations}, and finally verifies every memory access instruction in $R$ by classifying it into one of five categories and checking it with the simple and straightforward criteria of its belonging category as shown in Figure \ref{figure:verification:stage4}.

If a binary passes Stage 1-4 of the verification and is loaded into a domain by the LibOS, then we can prove the theorem below.

\begin{theorem}\label{theorem:memory_accesses}
Any memory access instruction in the domain is compliant with the memory access policy of MMDSFI (\S\ref{section:mmdsfi:overview}).
\end{theorem}

\begin{prf}
Any memory access instruction, regardless of how its destination is specified, has been either verified with the \CfiLabel-aware range analysis or rejected by our verification in Stage 4. Thus, we prove the theorem.
\end{prf}

\iffalse
\textbf{Stage 5 - Memory load verification.} This stage can be done the same way as the stage 4.
\fi
\section{Library OS}\label{section:libos}

In this section, we give an overview of the Occlum LibOS (see Figure \ref{figure:occlum:libos}) with an emphasis on its unique aspects.

\textbf{ELF loader.} A typical program loader in a Unix-like OS performs tasks such as parsing program binaries (e.g., ELFs), copying program images, and initializing CPU states. Beyond these basic tasks, the program loader in Occlum has four extra responsibilities.
First, the loader checks that the ELF binaries to be loaded are verified and signed by the Occlum verifier. 
Second, the loader rewrites all \CfiLabels in the program image so that the last four bytes of the \CfiLabels are set to the ID of the domain associated with the new SIP.
Third, the loader inserts into the process image a small piece of trampoline code that jumps to the entry point of the LibOS, thereby enabling the LibOS system calls. This trampoline code is the only way out of the sandbox enforced by MMDSFI. The address of this trampoline code is passed to libc via the auxiliary vector~\cite{AuxiliaryVector}.
Fourth, the loader initializes MPX bound registers according to the memory layout of the SIP's domain.
% TODO: admit that we only support statically-linked binaries. 

\textbf{Syscall interface.}
LibOS system calls are just function calls, except that the users must go through the trampoline code inserted by the ELF loader. With SIPs sandboxed by MMDSFI, the only chance for faulty SIPs to compromise the LibOS is through system calls. So, at the entry point of the LibOS is a piece of carefully-written assembly, which performs sanity checks, switches between user/LibOS stack, switches between user/LibOS thread-local storage, and eventually calls system call dispatching routine. After system call finishes, before returning to the SIP, LibOS will ensure that the return address target is a \CfiLabel of corresponding SIP.
%\textbf{ELF loader.} When invoked by \texttt{exec} syscall, a typical program loader in Unix-like OSes performs tasks like parsing program binaries (e.g., ELFs), copying program images, and initializing CPU states. Beyond these basic tasks, the program loader in Occlum has two extra responsibilities. First, the loader checks that the executables and shared libraries to be loaded are verified and signed by Occlum verifier. Second, the loader of Occlum LibOS also performs dynamic linking, which deviates from the UNIX tradition of running dynamic linker (\texttt{ld.so}) at the user level. Apparently, Occlum's moving dynamic linking into the LibOS is necessary since allowing an untrusted, user-level dynamic linker that can load arbitrary code into enclaves would break the MMDSFI-based isolation mechanism of Occlum LibOS.

\textbf{Memory management.}
Occlum preallocates the available enclave pages of an MMDSFI domain during enclave initialization. For each domain, the enclave pages are allocated and set with proper permissions according to the memory layout described in \S\ref{section:mmdsfi:overview}. The maxmimum size of the the code region is prespecified at compile time; so is the data region. The size of guard regions is set to 4KB. And the linker of the Occlum toolchain is aware of this 4KB gap between the code segment and the data segment when generating an ELF binary for Occlum.
The maximum number of MMDSFI domains is also prespecified at compile time.
Note that this preallocation of enclave pages is intended to work around the limitation of SGX 1.0 and can be avoided on SGX 2.0.

Memory mapping syscalls, e.g., \texttt{mmap} and \texttt{munmap}, can only manipulate the memory in the data region of a domain.
Anonymous memory mappings are fully supported, but requires the LibOS to manually initialize the allocated memory pages to zeros.
File-backed memory mappings is implemented by copying the file content to the mapped area.
Shared memory mappings are impossible because SGX cannot map one EPC page to multiple virtual addresses.
For the sake of security, the users are not allowed to add or remove the \texttt{X} permission of enclave pages (even on SGX 2.0). 

\textbf{Process management.} Occlum provides the \texttt{spawn} system call, instead of \texttt{fork}, to create SIPs. SIPs are mapped one-on-one to SGX threads, which are scheduled transparently by the host OS. This frees the LibOS from the extra complexity of process scheduling. IPC between SIPs, e.g., signals, pipes, and Unix domain sockets, are implemented efficiently via shared data structures in the LibOS.

LibOS threads are treated as SIPs that happen to share resources such as virtual memory, file tables, signal handlers, etc.
The synchronization between LibOS threads, e.g., \texttt{futex}, eventually relies on the host OS to sleep or wake up the corresponding SGX threads; but the semantic correctness of the synchronization primitives only relies on the LibOS.

\textbf{File systems.} To protect the confidentiality and integrity of persistent data, Occlum provides an encrypted file system that transparently encrypts all file data, metadata, and directories.
While Intel SGX SDK contains a library named Intel SGX Protected File System~\cite{SGXPFS}, this library can only protect the content of individual files, not file metadata and file directories. We build upon the primitives provided by this library to implement a complete encrypted file system. Occlum also provides special file systems (e.g., \texttt{/dev/}, \texttt{/proc/}, etc.), which are completely implemented by the LibOS inside the enclave. All SIPs share a common cache for file I/O. A child SIP can inherit the open file table of its parent SIP with minimal overhead. All SIPs have a unified view of the LibOS file systems.

\textbf{Networking.} Network-related operations are mostly delegated to the host OS, and the LibOS is only responsible for redirecting, bookkeeping, and sanity checks. So network I/O is not secure by default and requires user-level network encryption such as TLS for secure communication.
\section{Security Analysis}\label{section:security_analysis}

With all three components of Occlum described, we can now treat the Occlum system as a whole and examine whether we have achieved our overall security goal of inter-process isolation and process-LibOS isolation. 
More specifically, we will consider whether two common classes of attacks---code injection and return-oriented programming (ROP)---could be used by a malicious SIP to penetrate the isolation enforced by MMDSFI and the LibOS.

\textbf{Code injection attacks.}
SGX 1.0 does not allow enclave pages to be added, removed, or modified at runtime. To load programs into an enclave dynamically, the implementation of SGX LibOSes has to reserve a pool of enclave pages with \texttt{RWX} permissions when the enclave is launched. So the enclave is made susceptible to code injection attacks. This is a common pitfall of all existing SGX LibOSes.

Fortunately, Occlum is immune to code injection attacks. This is because 1) a SIP can only write to the data region of its domain, whose enclave pages have no executable permission (see \S\ref{section:mmdsfi:overview}); 2) only the LibOS can modify the executable enclave pages, e.g., when loading new verified binaries; and 3) the LibOS ensures system calls (e.g., \texttt{mmap} and \texttt{mprotect}) cannot be abused by SIPs.
In short, a malicious SIP cannot inject arbitrary code to bypass the isolation mechanism.

\textbf{ROP attacks.}
Now that a malicious SIP cannot inject new code, it can still attempt to reuse existing code gadgets for ROP attacks. ROP attacks are hard to prevent: it has been shown that static CFIs, including the coarse-grained CFI integrated with MMDSFI, cannot fully prevent ROP attacks~\cite{CFIBending,outofcontrol}.

But this does not change the fact that a malicious SIP cannot break the isolation enforced by MMDSFI and the LibOS. First of all, MMDSFI does make ROP attacks harder. The coarse-grained CFI enforced by MMDSFI prevents the untrusted code in a domain to jump to arbitrary locations; in fact, it can only jump to the \CfiLabels inside the domain. This greatly reduces the number of useful gadgets and thus the odds of success for ROP attacks. More importantly, any combination of code gadgets by ROP attacks has already been covered by the verifier (see \S\ref{section:verification}). So even ROP attacks cannot violate our security policies and break the isolation.
\section{Implementation}\label{section:implementation}

We have implemented the Occlum system, which consists of the toolchain, the verifier, and the LibOS. Our prototype implementation comprises over $20,000$ lines of source code in total ($15,000$ lines of code in Rust for LibOS, $3,000$ lines of code in C++ for toolchain, and $2000$ lines of code in Python for the verifier). The LibOS and the toolchain has been made open source on GitHub~\cite{OcclumGithub}. We describe briefly about how the three components are implemented.

\textbf{The Occlum toolchain.}
The toolchain is based on LLVM 7.0~\cite{LLVM}. In LLVM's x86 backend, we add two extra passes: one instruments control transfer instructions, and the other instruments memory access instructions and performs the range analysis-based optimizations.

We modify LLD so that the linker generates ELF executables that are compatible with MMDSFI. Specifically, the modified linker ensures that the generated code segments only contains code (no read-only data) and that a 4KB-gap between the code and data segments is reserved for the guard region between the code and data regions.

We modify musl libc~\cite{musl} to use the system calls provided by the Occlum LibOS. The \texttt{posix\_spawn} API is rewritten to use Occlum's \texttt{spawn} instead of \texttt{vfork} and \texttt{execve}.% And the modified musl libc is also compiled with the Occlum toolchain. All ELF executables are linked with this modified musl libc by default.

\textbf{The Occlum verifier.}
The implementation of the verifier depends on two libraries. In order to decode individual x86-64 instructions correctly, we import Zydis~\cite{Zydis}, a disassembler library supporting all x86-64 instructions, including all the SGX and MPX instructions we need. In addition, we import PyVEX~\cite{PyVEX}, a library that converts x86-64 instructions into its VEX intermediate representation (IR). This reduces the task of understanding the complex semantics of x86-64 instructions into manipulating the simple VEX IR instructions.% The codebases of the verifier and the toolchain are exclusive to each other. So bugs in the toolchain are unlikely to appear in the verifier.

\textbf{The Occlum LibOS.}
The LibOS is mostly written in Rust~\cite{Rust}, a memory-safe programming language. Thanks to Rust, we can minimize the odds of low-level, memory-safety bugs in the LibOS.
The LibOS is based on Intel SGX SDK~\cite{SGXSDK} and Rust SGX SDK~\cite{RustSGXSDK} for the in-enclave runtime support of C and Rust, respectively.
In addition, we implement a FUSE~\cite{FUSE}-based utility to mount and manipulate Occlum's encrypted file system in the host environment. This simplifies the task of preparing encrypted FS images for the Occlum LibOS in development environments.
Currently, the LibOS only supports loading statically-linked ELF executables; we will add support for shared libraries in the future. %We also write a simple Linux kernel module to enable \texttt{wrfsbase} instruction at user level. 
\section{Evaluation}\label{section:evaluation}

In this section, we present the experimental results that answer the following questions:
(1) To what extent can Occlum improve the overall performance of real-world, multi-process applications? (\S\ref{section:evaluation:application_benchmarks})
 (2) What about the performance of individual system calls? (\S\ref{section:evaluation:system_call_benchmarks})
(3) Does the use of MMDSFI make user applications more secure? And how much overheads does MMDSFI incur?(\S\ref{section:evaluation:mmdsfi_benchmarks})

To answer the above questions, we compare Occlum with Linux and the state-of-the-art SGX LibOS, Graphene-SGX.
The time of process creation on Graphene-SGX is sensitive to the sizes of the enclaves. For any benchmark whose result is affected by the time of process creation, we configure Graphene-SGX to use the minimal enclave size that is able to run the benchmark.
\iffalse
To reveal the impact of SFI on system performance, applications in Occlum use Occlum-none that without instrumentation, Occlum-int that constrain only writes and Occlum-conf which constrain both reads and writes. Note our library OS security features are configured as disable for testing.
\fi
\iffalse
Unless explicitly stated, all ELF executables in Occlum benchmarks are assumed to be benign-yet-faulty, thus instrumented with only the integrity protection (i.e., no confidentiality protection).
\fi

% TODO: the setup needs confirmation from Youren
\textbf{Experimental setup.} We use machines with a two-core, 3.5GHz Intel Core i7 CPU (hyper-threading disabled), 32GB memory, and a 1TB SSD disk, and 1Gbps Ethernet card. Each machine supports SGX 1.0. The host runs Linux kernel 4.15. We install Intel SGX SDK v2.4.0, Rust nightly-2019-01-28, and Rust SGX SDK v1.0.6. We use the latest version of Graphene-SGX (commit f30f7e7) at the time of experiment.

\subsection{Application benchmarks}\label{section:evaluation:application_benchmarks}

\begin{figure*}[t]\label{figure:evaluation:applications}
\newcommand{\myHeight}{3.7cm}

\begin{subfigure}[t]{0.3\textwidth}
    \centering
    \includegraphics[height=\myHeight]{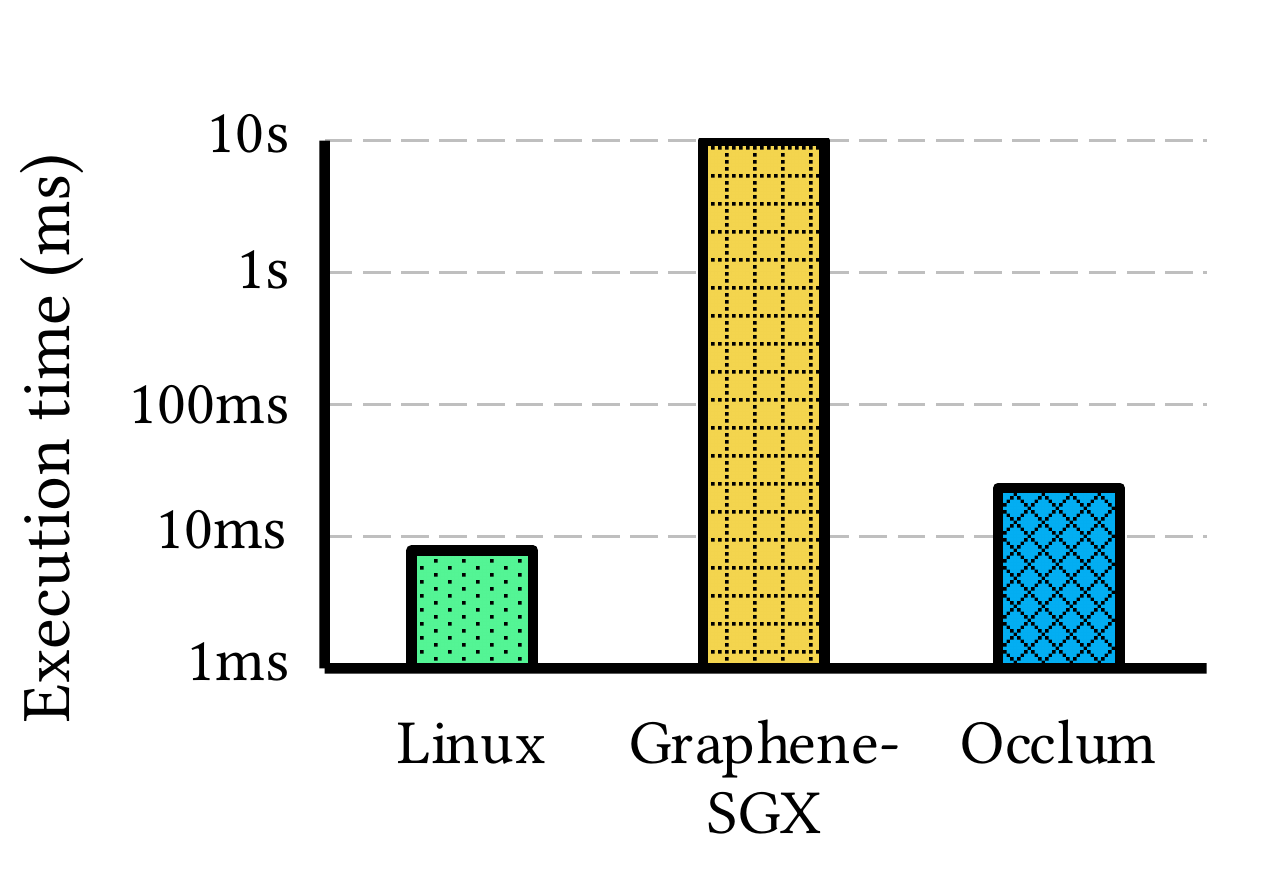}
    \caption{Fish}
    \label{figure:evaluation:fish}
\end{subfigure}
~ 
\begin{subfigure}[t]{0.35\textwidth}
    \centering
    \includegraphics[height=\myHeight]{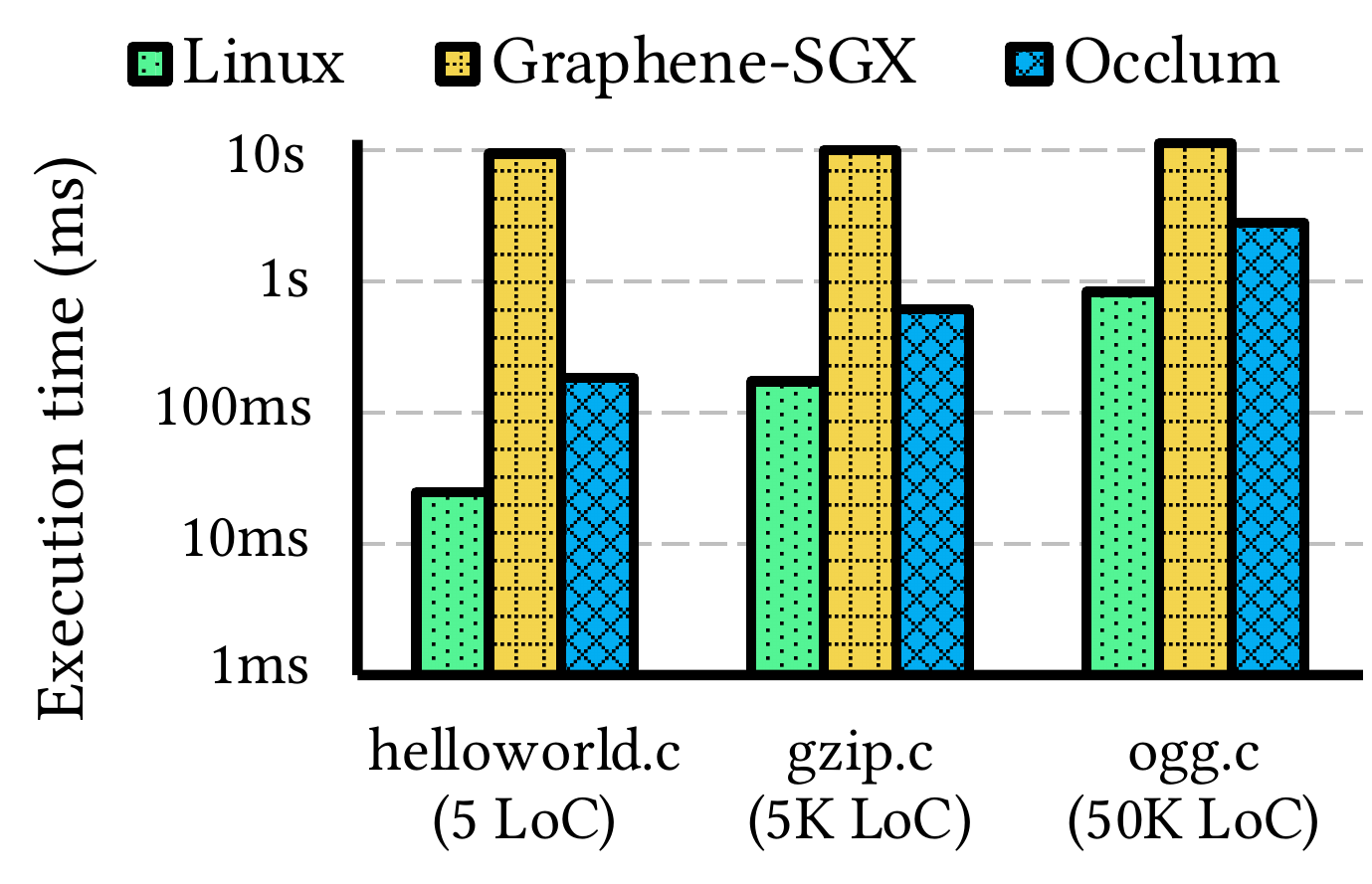}
    \caption{GCC}
    \label{figure:evaluation:gcc}
\end{subfigure}
~ 
\begin{subfigure}[t]{0.35\textwidth}
    \centering
    \includegraphics[height=\myHeight]{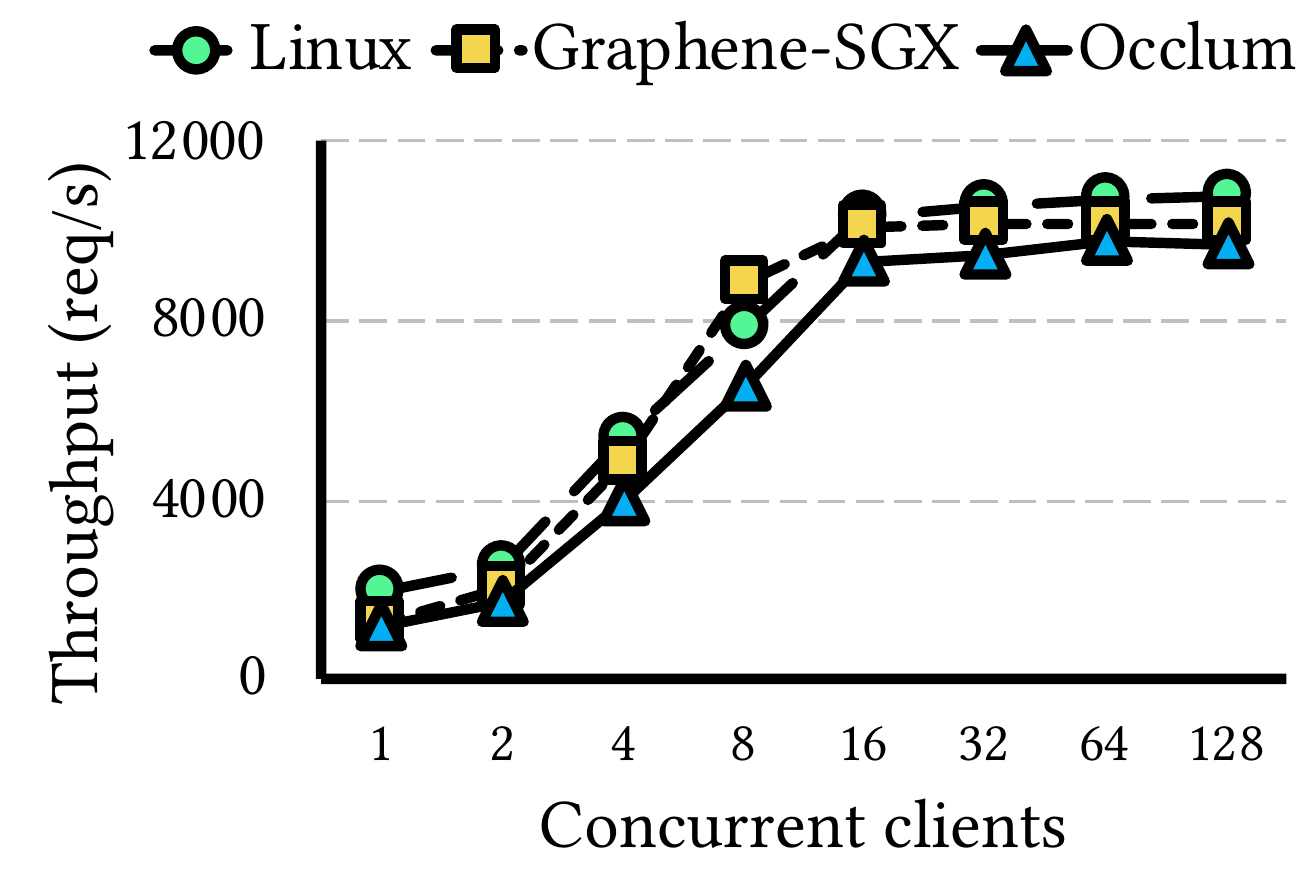}
    \caption{Lighttpd}
    \label{figure:evaluation:lighttpd}
\end{subfigure}

\caption{Application benchmarks.}

\let\myHeight\undefined
\end{figure*}

We measure the performance of three widely-used applications:
\textbf{Fish} (v3.0.0), a user-friendly command line shell~\cite{Fish};
\textbf{GCC} (v4.4.5), the GNU compiler for C-family programming languages~\cite{Gcc}; and
\textbf{Lighttpd} (v1.4.40), a fast Web server~\cite{Lighttpd}.
All of them are multi-process applications but
have distinct workload characteristics:
Fish is \emph{process}-intensive (as every shell command is executed in a separate process),
GCC is \emph{CPU}-intensive, and
Lighttpd is \emph{I/O}-intensive.

\textbf{Fish.}
We use the shell script provided by UnixBench~\cite{UnixBench}.
The test script applies on the data from an input file a series of transformation using multiple utilities (e.g., sort, od, and grep), which are connected via pipes or I/O redirections.
Fish does not need any code modification to run on Occlum since it uses the modern \texttt{posix\_spawn}, rather than \texttt{fork}, to create processes. 
As shown in Figure \ref{figure:evaluation:fish}, Occlum (19.5ms) is $13.9\times$ slower than Linux (1.4ms), but nearly $500\times$ faster than Graphene-SGX (9.5s). Occlum is slower than Linux due to the lack of on-demand loading in enclave (see \S\ref{section:evaluation:system_call_benchmarks}).
%The performance differences stem from the process startup and IPC (see \S\ref{section:evaluation:system_call_benchmarks}).

\textbf{GCC.}
We use three C files of varied sizes: the first one is a ``Hello World!'' program (5 LOC), the other two are real-world programs (5K and 50K LOC, respectively) collected by MIT~\cite{MIT}.
While GCC is CPU-intensive in nature, it needs to create separate processes for its preprocessor, compiler, assembler, and linker. 
To start these child processes, GCC originally uses \texttt{fork}; we refactor the source code to use \texttt{posix\_spawn} instead, which is is about 50 LOC.
As shown in Figure \ref{figure:evaluation:gcc}, the compilation time of the three C files ranges from 25ms to 830ms on Linux, from 9.7s to 11.7s on Graphene-SGX, and from 229ms to 3.0s on Occlum.
In other words, Occlum is $3.6\times - 9.2\times$ slower than Linux, but $3.82 \times - 42\times$ faster than Graphene-SGX.

\textbf{Lighttpd.}
We use ApacheBench~\cite{ApacheBench} on a client machine to generate HTTP requests that retrieve 10KB HTML pages from an instance of Lighttpd Web server running on Linux, Graphene-SGX, or Occlum. Both the client and server are in the same local area network. We gradually increase the concurrency of ApacheBench to simulate an increasing number of concurrent clients.
We configure the master process of Lighttpd to start two worker processes, both of which inherit the listening sockets from the master and handle the connection requests to the listening sockets together. Similar to GCC, we refactor the source code to replace \texttt{fork} with \texttt{posix\_spawn}, in about 150 LOC. The results of Lighttpd are shown in Figure \ref{figure:evaluation:lighttpd}. The peak throughput of both Graphene-SGX ($10\%$ overhead) and Occlum ($9\%$ overhead) is slightly lower than that of Linux. 
%We suspect the extra overhead of Occlum compared to Graphene-SGX stems from some differences in implementation, not in design. We will further investigate the issue and do more optimizations in the future.

The results above show that Occlum can deliver a significant performance boost to some multi-process applications.% And refactoring legacy code to replace \texttt{fork} with \texttt{spawn} is practical.

\subsection{System call benchmarks}\label{section:evaluation:system_call_benchmarks}

\begin{figure*}[t]\label{figure:evaluation:system_call}
\newcommand{\myHeight}{3.8cm}

\begin{subfigure}[t]{0.24\textwidth}
    \centering
    \includegraphics[height=\myHeight]{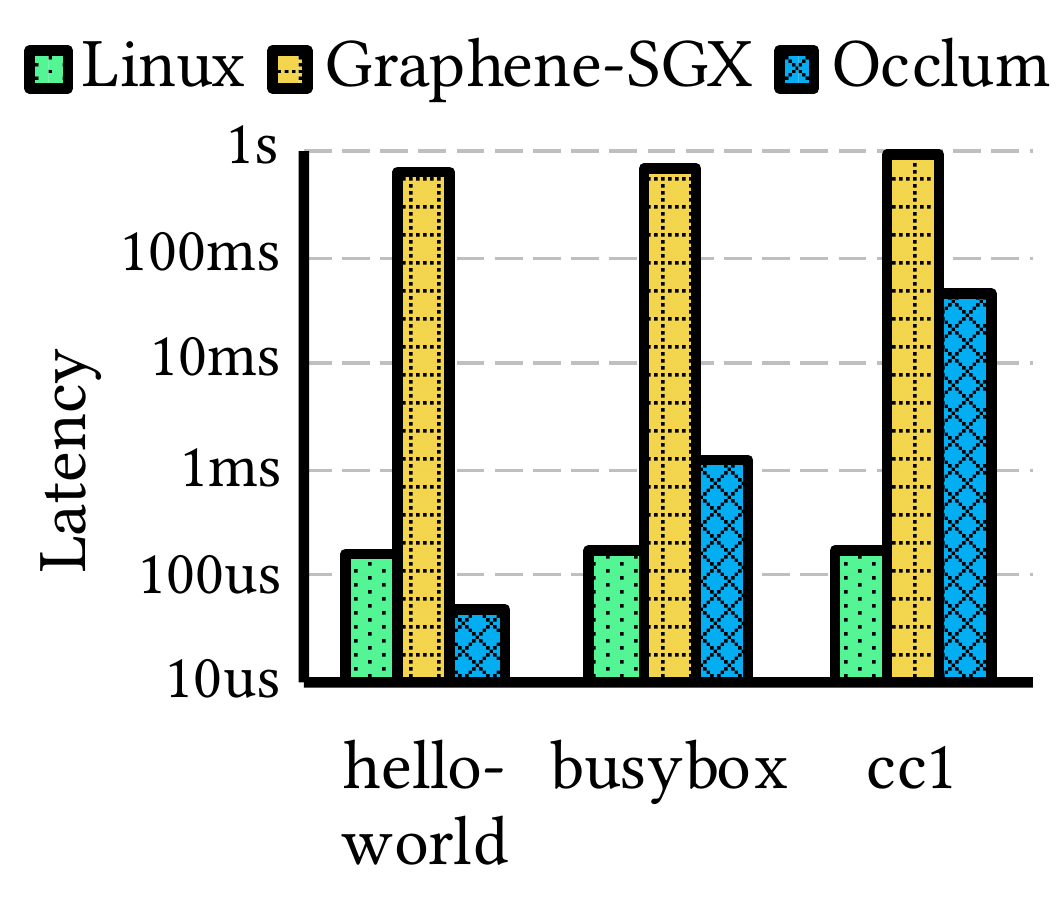}
    \caption{Process creation}
    \label{figure:evaluation:spawn}
\end{subfigure}
~ 
\begin{subfigure}[t]{0.24\textwidth}
    \centering
    \includegraphics[height=\myHeight]{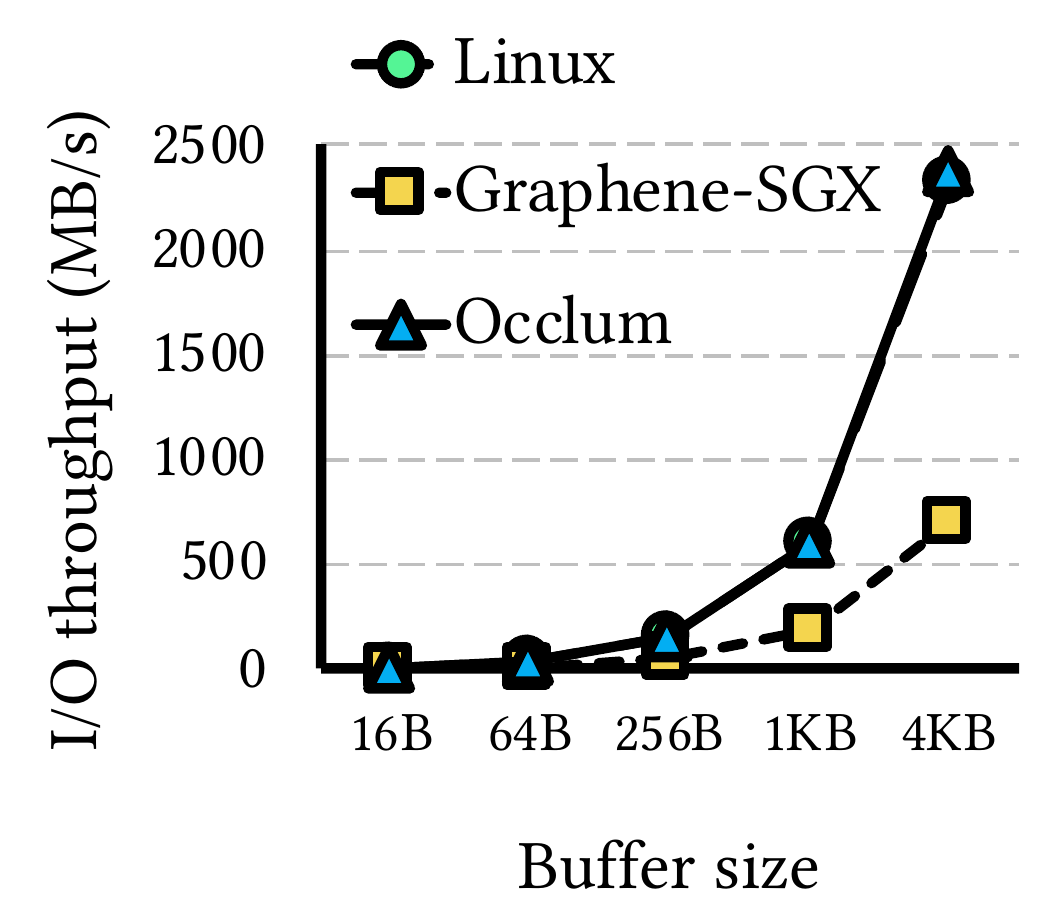}
    \caption{Pipe}
    \label{figure:evaluation:pipe}
\end{subfigure}
~ 
\begin{subfigure}[t]{0.24\textwidth}
    \centering
    \includegraphics[height=\myHeight]{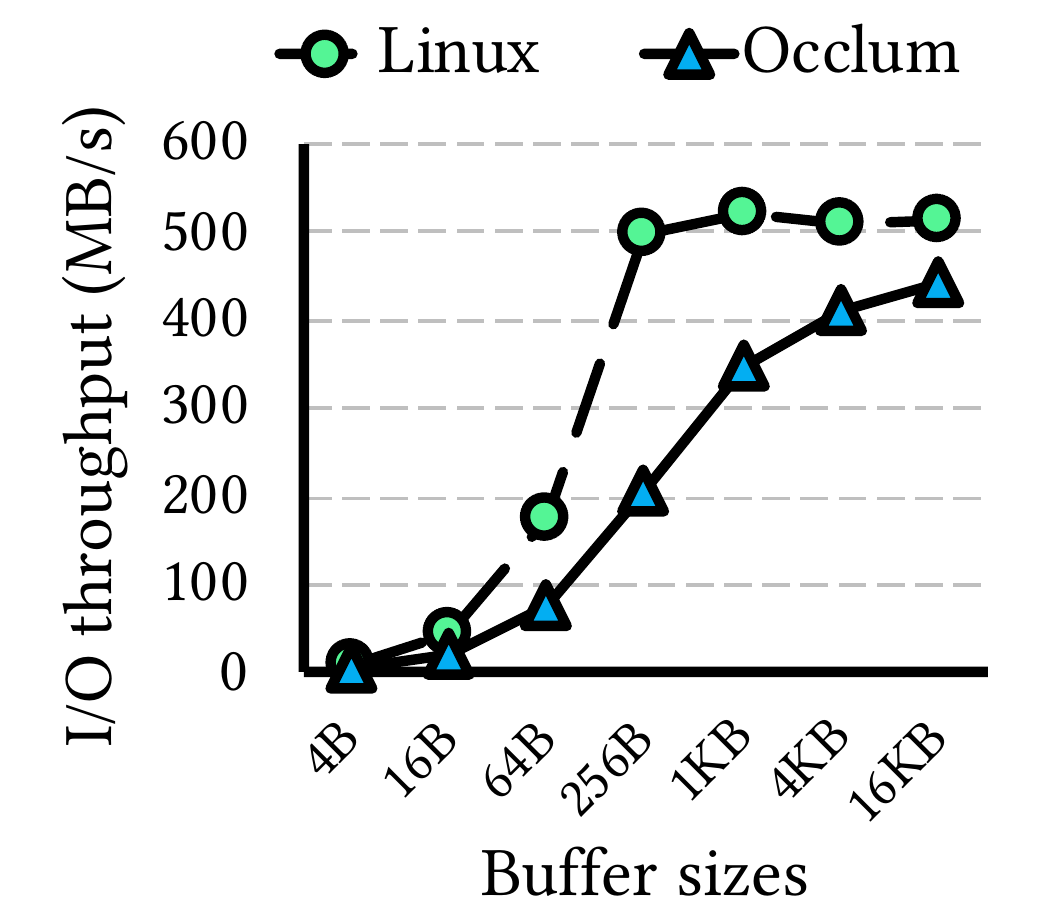}
    \caption{File reads}
    \label{figure:evaluation:file_reads}
\end{subfigure}
~ 
\begin{subfigure}[t]{0.24\textwidth}
    \centering
    \includegraphics[height=\myHeight]{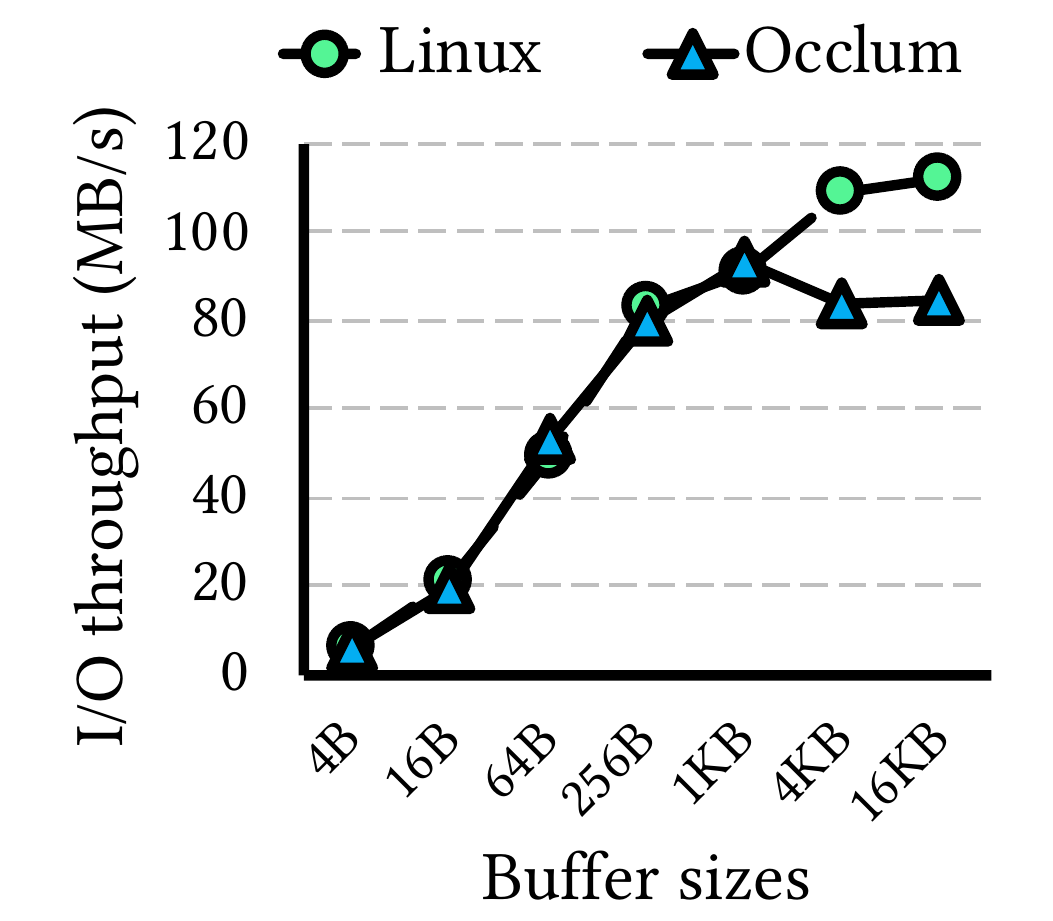}
    \caption{File writes}
    \label{figure:evaluation:file_writes}
\end{subfigure}

\caption{System call benchmarks.}

\let\myHeight\undefined
\end{figure*}

To better understand the results of the application benchmarks above, we measure the performance of some individual system calls. This helps us identify the sources of performance gain or loss due to the Occlum LibOS.

\textbf{Process creation time.}
We measure the process creation time of different binaries on Linux, Graphene-SGX, and Occlum. 
There are three binaries of different sizes: 1) a ``Hello World" program of 14KB size, 2) busybox, a collection of common UNIX utilities combined into a single small executable of 400KB size, and 3) cc1, the GCC front-end compiler of 14MB size.
%One small test uses a ``Hello World!" program with the size of 14KB. One median test uses the busybox program with the size of 400KB. And one large test uses the GCC front-end compiler, cc1 with the size of 14MB. 
%We control the running of the median and large tests to exit the main function immediately by setting the input parameters. 
Process are created using libc's \texttt{posix\_spawn}.
Occlum implements \texttt{posix\_spawn} using its spawn system call. Linux and Graphene-SGX use \texttt{vfork} and \texttt{execve} to implement \texttt{posix\_spawn}. \texttt{vfork} is more efficient than \texttt{fork} as it avoids copying the page tables on Linux and copying the process state on Graphene-SGX.
%We configure Graphene-SGX to use 16MB enclaves to run the small and median tests as 16MB is the minimal requirement for starting an EIP. The large test uses 256MB enclaves because of the binary size and other memory consumption. 
%As in \ref{figure:evaluation:spawn}, The creation time of all program are 170 us on Linux, from 97us to 63ms on Occlum and from 0.64s to 0.89s on Graphene-SGX. That is say, Occlum are faster than graphene-SGX from $13\times$ to $6,600 \times$.
As shown in \ref{figure:evaluation:spawn}, Linux consumes about $170$us regardless of the specific binary. For the small-sized binary ($14$KB), Occlum consumes $97$us, which is $1.6\times$ faster than Linux and over $6,600 \times$ faster than Graphene-SGX ($0.64$s). For the median-sized binary, Occlum consumes $1.7$ms, which is $10.5\times$ slower than Linux, but over $390\times$ faster than Graphene-SGX ($0.69$s). For the large-sized binary, Occlum consumes $63$ms, which is $13\times$ faster than Graphene-SGX ($0.89$s).
%Regardless of the binary sizes,% when creating a process, 
Linux initializes a process's page table without actually loading all the pages with their data from disks. Thus, the process creation time on Linux is insensitive to binary sizes. Occlum has to load the entire binaries into the enclave. So the process creation time on Occlum is proportional to the sizes of binaries. Graphene-SGX has to create a new enclave for each new process, which is very time consuming.

%We measure the latency to create a minimal process by using libc's  on Linux, Graphene-SGX, and Occlum. On Occlum,  is directly implemented in Occlum's \texttt{spawn} system call; on Linux and Graphene-SGX, it is implemented in \texttt{vfork} and \texttt{execve}. \texttt{vfork} is more efficient than \texttt{fork} as it avoids copying the page tables on Linux and copying the process state on Graphene-SGX. We configure Graphene-SGX to use 16MB enclaves as 16MB is the minimal enclave size that is possible start an EIP. The binary to be run in the process is a ``Hello World!'' program. As shown in Figure \ref{figure:evaluation:spawn}, the latency of the process creation is $0.17$ms on Linux, $0.86$s on Graphene-SGX, $0.51$ms on Occlum. In other words, Occlum is $3.1\times$ slower than Linux, but over $1,600\times$ faster than Graphene-SGX. Occlum is slower than Linux because (1) Occlum loads binaries from the encrypted file systems, which involves the extra overhead of decryption; and (2) Occlum cannot use the copy-on-write technique inside enclaves to allocate empty pages and must fill pages with zeros manually, which takes more time.

\textbf{IPC throughput.}
We measure the throughput of IPC by using a common IPC method, pipe. We create two processes that are connected via a pipe. The throughput of the pipe is measured under varied buffer sizes.
As shown in Figure \ref{figure:evaluation:pipe}, Occlum's throughput is on par with Linux's throughput, which is over $3\times$ higher than Graphene-SGX's throughput.

\textbf{File I/O throughput.}
To quantify the performance overhead of Occlum's transparent file encryption, we compare Occlum's encrypted file system with Linux's Ext4. And we exclude Graphene-SGX from this benchmark since Graphene-SGX does not have a fully-fledged encrypted file system.
The throughput of sequential file reads and writes under different buffer sizes are shown in Figure \ref{figure:evaluation:file_reads} and \ref{figure:evaluation:file_writes}.
Compared with Ext4, Occlum incurs an average overhead of $39\%$  on file reads and an average overhead of $18\%$ on file writes. 

The above results show that
the Occlum LibOS greatly improves the performance of process startup and IPC, while moderately degrades the performance of file I/O in exchange for the transparent encryption of persistent data.

\subsection{MMDSFI benchmarks}\label{section:evaluation:mmdsfi_benchmarks}

\begin{figure}[t!]\label{figure:evaluation}
\newcommand{\myHeight}{4.5cm}

\begin{subfigure}[t]{0.5\columnwidth}
    \centering
    \includegraphics[height=\myHeight]{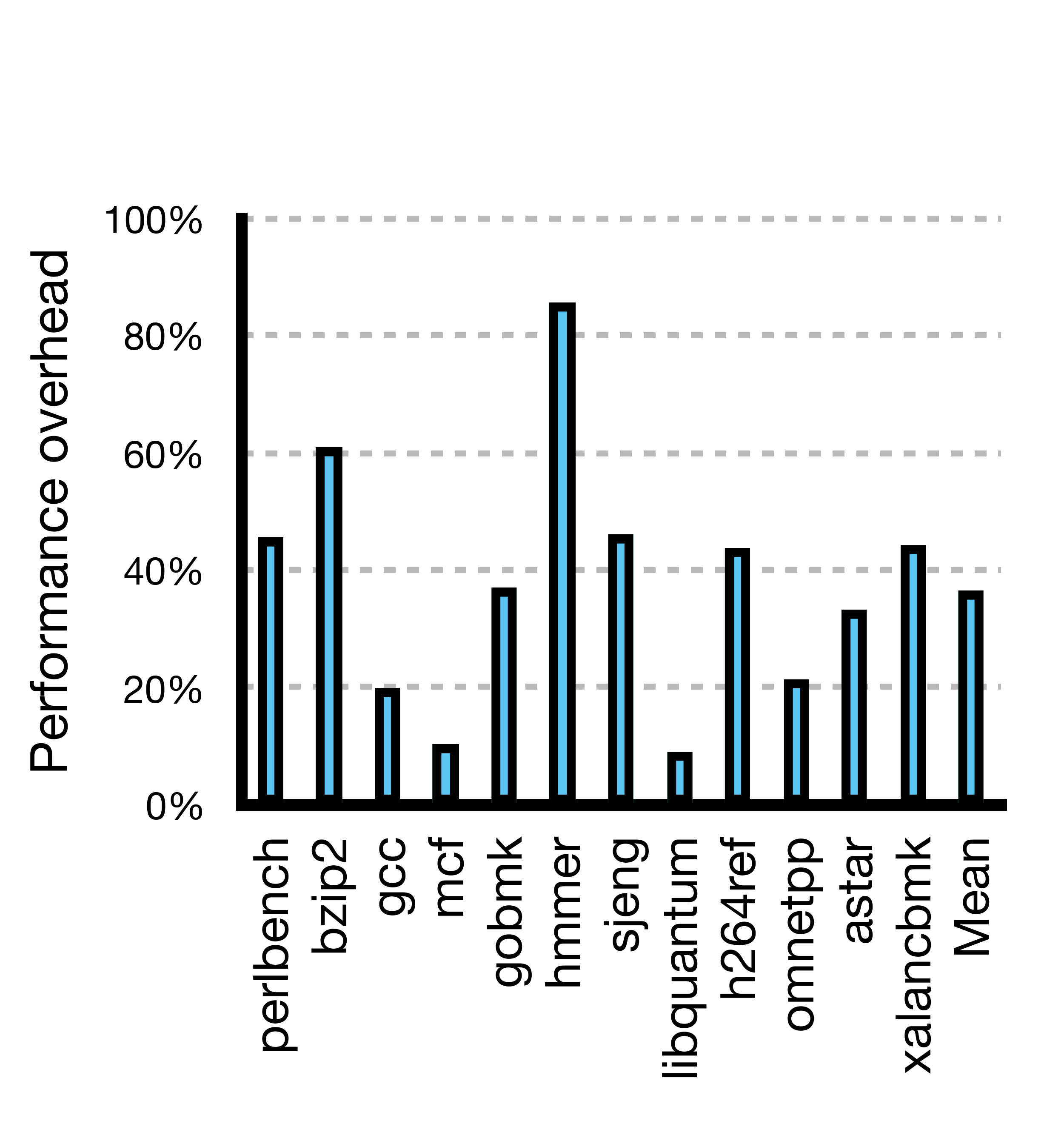}
    \caption{The overheads on SPECint2006}
    \label{figure:evaluation:spec_benchmarks}
\end{subfigure}
~
\begin{subfigure}[t]{0.5\columnwidth}
    \centering
    \includegraphics[height=\myHeight]{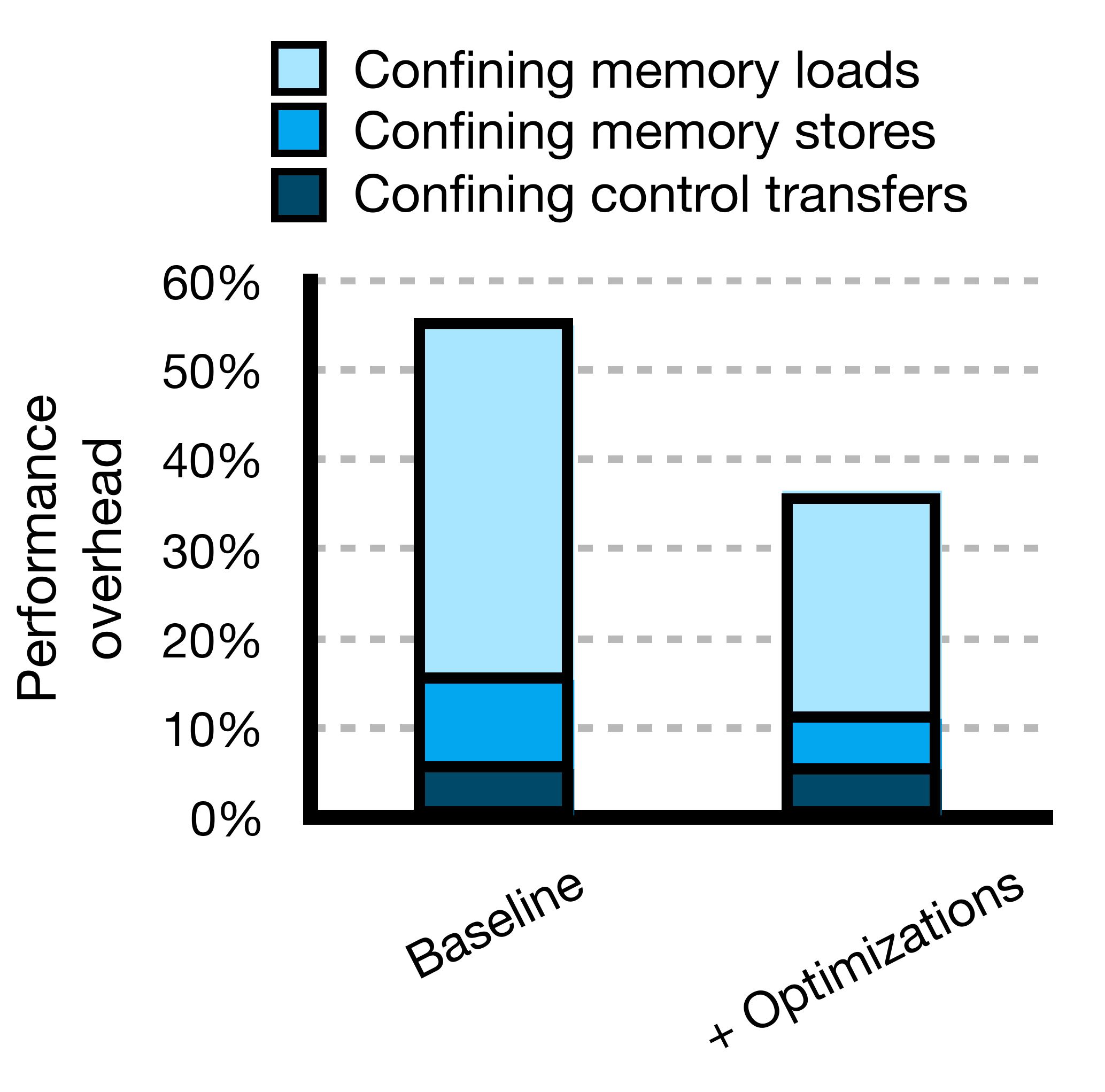}
    \caption{The overhead breakdown}
    \label{figure:evaluation:overhead_breakdown}
\end{subfigure}

\caption{MMDSFI benchmarks}

\let\myHeight\undefined
\end{figure}

\textbf{Security.}
To demonstrate the effectiveness of MMDSFI, we use RIPE~\cite{RIPE} benchmark, which is designed to compare different defenses against attacks that exploit buffer-overflow bugs.
RIPE builds up a total of 850 workable attacks, including code injection, ROP, return-to-libc, etc. We run RIPE on Graphene-SGX and Occlum for comparison.

When the stack protection is disabled by the compiler, $36$ code injection, $2$ ROP, and $16$ return-to-libc attacks succeed on Graphene-SGX. When the stack protection is enabled, the successful attacks are reduced to $16$ code injection and $12$ return-to-libc.
With or without the stack protection, Occlum can prevent all code injection and ROP attacks in RIPE. ROP attacks in RIPE can be fully prevented by MMDSFI as the ROP gadgets in RIPE do not start with \CfiLabels.
Return-to-libc attacks still succeed ($16$ without stack protection and $12$ with stack protection) since libc functions are normal functions starting with \CfiLabels.
These results show that MMDSFI is effective in defending against memory-related attacks. Note that the successful attacks (e.g., return-to-libc) that are not prevented by MMDSFI do not break the isolation between SIPs.

\textbf{CPU overhead.}
While MMDSFI is intended for the use in the Occlum LibOS, it can be potentially applied in other use cases. So it is good to know the overheads of MMDSFI alone, independent from the Occlum LibOS. 
We use the SPECint2006~\cite{CINT2006} to measure the overheads of MMDSFI on CPU-intensive workloads. As shown in Figure \ref{figure:evaluation:spec_benchmarks}, the average overhead of MMDSFI is $36.6\%$.

\textbf{The breakdown of CPU overheads.}
To further understand the overheads of MMDSFI, we break down the overheads into three sources: confining control transfers, confining memory stores, and confining memory loads. And we conduct this breakdown analysis on two implementations of MMDSFI: (1) the naive implementation that inserts \MemGuards for all memory loads and stores; (2) the optimized implementation that leverages the range analysis-based optimization to eliminate unnecessary \MemGuards (\S\ref{section:mmdsfi:optimizations}). Figure \ref{figure:evaluation:overhead_breakdown} show that the range analysis-based optimizations are effective, reducing the overhead of confining memory stores from $10.1\%$ to $4.3\%$ and the overhead of confining memory loads from $39.6\%$ to $25.5\%$.
%\input{sections/discussion}
\iffalse
\section{Discussions}\label{section:discussions}

Support JIT code

\fi

\section{Conclusions}\label{section:conclusion}

In this paper, we present Occlum, a system that enables secure and efficient multitasking on SGX LibOSes. We implement the LibOS processes as \emph{SFI-Isolated Processes (SIPs)}. To this end, we propose a novel SFI scheme named \emph{MPX-based, Multi-Domain SFI (MMDSFI)}. We also design an independent verifier to ensure the security guarantees of MMDSFI. With SIPs safely sharing the single address space of an enclave, the LibOS can implement multitasking efficiently. Experimental results show that Occlum outperforms the state-of-the-art multitasking SGX LibOS significantly.
\begin{acks}\label{section:ack}

We would like to thank ATC, SOSP and ASPLOS anonymous reviewers for their helpful comments. And we would also like to thank Junjie Mao for his valuable insights.

The authors from Tsinghua University are supported by National Key R\&D Program of China (Grant No. 2017YFB0802901), National Natural Science Foundation of China (Grant No. 61772303 and 61877035), Young Scientists Fund of the National Natural Science Foundation of China (Grant No. 61802219). The author from Shanghai Jiao Tong University is sponsored by National Natural Science Foundation of China (Grant No. 61972244) and Program of Shanghai Academic/Technology Research Leader (No. 19XD1401700). Some authors are sponsored by Beijing National Research Center for Information Science and Technology (BNRist) to attend the conference.
\end{acks}

\bibliographystyle{ACM-Reference-Format}
\balance
\bibliography{paper}

\newpage
\appendix

\section{Artifact Appendix}

\subsection{Abstract}

Our artifacts consist of the Occlum toolchain, the Occlum LibOS, and some scripts to set up the environment, build the projects, and run the benchmarks. The hardware requirement is Intel x86-64 CPUs with SGX and MPX support.

\subsection{Artifact check-list (meta-information)}

{\small
\begin{itemize}
%  \item {\bf Algorithm: }
%  \item {\bf Program: SPEC CPU 2006}
  \item {\bf Compilation: } LLVM (which will be downloaded and patched by the script).
  \item {\bf Transformations: } MMDSFI instrumentation implemented as LLVM passes.
 % \item {\bf Binary: }
 % \item {\bf Data set: }
  \item {\bf Run-time environment: } Root access to Ubuntu Linux.
  \item {\bf Hardware:  } Intel x86-64 CPU with SGX and MPX extensions. We recommend i7 Kaby Lake CPUs. To test network performance, another computer is required to connect as client with 1Gbps ethernet.
%  \item {\bf Run-time state: }
%  \item {\bf Execution: }
%  \item {\bf Metrics: }
  \item {\bf Output:  } For benchmarks, the results are printed in consoles.
  \item {\bf Experiments:  } Using Bash scripts.
  \item {\bf How much disk space required (approximately)?: } 40GB.
  \item {\bf How much time is needed to prepare workflow (approximately)?: } Half a day.
  \item {\bf How much time is needed to complete experiments (approximately)?: } One hour.
  \item {\bf Publicly available?: } Yes.
  \item {\bf Code licenses (if publicly available)?: } BSD.
 % \item {\bf Data licenses (if publicly available)?: }
 % \item {\bf Workflow framework used?: }
  \item {\bf Archived (provide DOI)?: } 10.5281/zenodo.3565239
\end{itemize}
}

%%%%%%%%%%%%%%%%%%%%%%%%%%%%%%%%%%%%%%%%%%%%%%%%%%%%%%%%%%%%%%%%%%%%%
\subsection{Description}

\subsubsection{How delivered}

The artifacts are available on GitHub: \url{https://github.com/occlum/reproduce-asplos20}. We need to modify source code from other projects like LLVM. Our scripts will automatically download the original source code and then apply the patches.

Note that the artifacts provided are the archive of an early version of Occlum project. We recommend interesting users to test on the latest version of Occlum, which can be found at \url{https://github.com/occlum/occlum}.

\subsubsection{Hardware dependencies}
We recommend testing on an Intel Core i7 Kaby Lake CPU (i7-7567U was used in our test). SGX and MPX are required.

\subsubsection{Software dependencies}
We built and tested our system on Ubuntu 16.04 with kernel 4.15.0-65-generic.

%\subsubsection{Data sets}

%%%%%%%%%%%%%%%%%%%%%%%%%%%%%%%%%%%%%%%%%%%%%%%%%%%%%%%%%%%%%%%%%%%%%
\subsection{Installation}
You can get the artifacts from GitHub using the following command:
\begin{verbatim}
git clone https://github.com/occlum/reproduce-asplos20
\end{verbatim}

%%%%%%%%%%%%%%%%%%%%%%%%%%%%%%%%%%%%%%%%%%%%%%%%%%%%%%%%%%%%%%%%%%%%%
\subsection{Experiment workflow}

The overall workflow consists of the following steps:
\begin{enumerate}
    \item Install the dependencies;
    \item Build the LibOS;
    \item Build the toolchain;
    \item Build and run the macro-benchmarks;
    \item Build and run the micro-benchmarks.
\end{enumerate}

We provide scripts for each of the steps above.

%%%%%%%%%%%%%%%%%%%%%%%%%%%%%%%%%%%%%%%%%%%%%%%%%%%%%%%%%%%%%%%%%%%%%
\subsection{Evaluation and expected result}

We will go through the entire experiment workflow by describing all the commands in each step.

\subsubsection{Install the dependencies} Run the following command:
\begin{verbatim}
./prepare.sh
\end{verbatim}

\subsubsection{Build the LibOS} Run the following command:
\begin{verbatim}
./download_and_build_libos.sh.
\end{verbatim}
This script above will download the source code of the LibOS and build it.

\subsubsection{Build the toolchain.} To build and install the toolchain, run the following command:
\begin{verbatim}
cd toolchain && \
  ./download_and_build_toolchain.sh
\end{verbatim}
This script will install the toolchain at \texttt{/usr/local/occlum}. To use this toolchain for building benchmarks, export the installation directory to \texttt{PATH} with the following command:
\begin{verbatim}
export PATH=/usr/local/occlum/bin:$PATH
\end{verbatim}

\subsubsection{Build and run the macro-benchmarks.} As described in \S\ref{section:evaluation}, there are three macro-benchmarks: fish, GCC, and lighttpd.

\textbf{fish benchmark.} To build the benchmark, run the following command:
\begin{verbatim}
cd apps/fish && ./download_and_build_fish.sh
\end{verbatim}
After finsih building the benchmark, run the following command to run the benchmark:
\begin{verbatim}
cd apps/fish && ./run_fish_fish.sh
\end{verbatim}
When the script is done, the LibOS will print the total running time of the benchmark repeating for $100$ times.

\textbf{GCC benchmark.} To build GCC with the Occlum toolchain, run the following command:
\begin{verbatim}
cd apps/gcc && ./download_and_build_gcc.sh
\end{verbatim}
There are three workloads, which can be tested with the following commands:
\begin{verbatim}
cd apps/gcc && \
  ./run_gcc_helloworld.sh && \
  ./run_gcc_5K.sh && \
  ./run_gcc_50K.sh
\end{verbatim}

\textbf{lighttpd benchmark.}
To build lighttpd, run the following command:
\begin{verbatim}
cd apps/lighttpd && ./download_and_build_lighttpd.sh
\end{verbatim}

Next, open \texttt{apps/lighttpd/config/lighttpd-server.conf} with your favourite text editor and set \texttt{server.bind} to the IP address of your machine.

There are two modes of the benchmark. Start a single-thread server with 
\begin{verbatim}
cd apps/lighttpd &&
  ./run_lighttpd_test.sh
\end{verbatim}
or start a multi-threaded server with
\begin{verbatim}
cd apps/lighttpd &&
  ./run_lighttpd_test_multithread.sh
\end{verbatim}

After the server started, you can now test the throughput and latency by starting the http benchmark program:
\begin{verbatim}
cd apps/lighttpd &&
  ./benchmark-http.sh ${SERVER_IP}:8000
\end{verbatim}
where \texttt{SERVER\_IP} is the IP address of the lighttpd server. This benchmark will show the latency and throughput under different number of concurrent clients.

\subsubsection{Build and test micro-benchmarks} There are two micro-benchmarks: one measures the latency of process creation and the other measures the throughput of IPC.

\textbf{Spawn benchmark.} Run the benchmark with the following command:
\begin{verbatim}
cd bench/spawn && ./run_spawn_bench.sh
\end{verbatim}
The result will be printed on the console.

\textbf{Pipe benchmark.} Run the benchmark with the following command:
\begin{verbatim}
cd bench/pipe && ./run_pipe_bench.sh
\end{verbatim}
The result will be printed on the console.

\subsection{Experiment customization}
Since the LibOS and the toolchain are installed, users can build their own applications with our toolchain and run them with our LibOS.

\end{document}